\documentclass[aps,prd,reprint,amsmath,amssymb,superscriptaddress,floatfix]{revtex4}
\usepackage{graphicx}
\usepackage{amsfonts}
\usepackage{amssymb}
\usepackage{rotating}
\usepackage{booktabs}
\usepackage{xcolor}
\usepackage{soul}
\usepackage{color}
\usepackage{slashed}
\usepackage{multirow}
\usepackage{makecell}
\usepackage{epsf}
\usepackage{ulem}
\usepackage{cancel}
\usepackage{color,bm}

\usepackage{subfig}
\usepackage[justification=raggedright]{caption}

\usepackage[colorinlistoftodos]{todonotes}

\usepackage[colorlinks=true,citecolor=cyan,urlcolor=blue,bookmarks=true,bookmarks=true,bookmarksopen=true,bookmarksnumbered=true,bookmarksopenlevel=3]{hyperref}

\definecolor{airforceblue}{rgb}{0.36, 0.54, 0.66}
\definecolor{steelblue}{rgb}{0.27, 0.51, 0.71}
\definecolor{amber}{rgb}{1.0, 0.49, 0.0}

\def\comment#1{}

\begin{document}

\title{Double quarkonium hadroproduction as a probe of gluon Sivers function}
\date{\today}
\author{\textsc{Xuan Luo}}
\author{\textsc{Hao Sun}\footnote{Corresponding author: haosun@mail.ustc.edu.cn \hspace{0.2cm} haosun@dlut.edu.cn}}
\author{\textsc{Tichouk}}
\affiliation{Institute of Theoretical Physics, School of Physics, Dalian University of Technology, \\ No.2 Linggong Road, Dalian, Liaoning, 116024, P.R.China }

\begin{abstract}
The prediction of single spin asymmetry (SSA) in double $\rm J/\psi$ production in proton-proton collision is given out within the framework of non-relativistic QCD, using the recently obtained best fit parameters for the gluon Sivers function extracted from PHENIX data in $\rm p+p^\uparrow\to\pi^0+X$. The color singlet state $\rm ^3 S_1^{(1)}$ and color octet state $\rm ^3 S_1^{(8)}$ are considered to the SSA contribution in the double $\rm J/\psi$ hadroproduction. Our result shows that a sizable asymmetry can be estimated as functions of different kinematic variables.
\end{abstract}
\maketitle

\section{INTRODUCTION}
\label{introduction}

Since it was first observed, single spin asymmetry (SSA) is a topic in spin physics of significant theoretical and experimental interest \cite{Adams:1991rw,Adams:1991cs,Arsene:2008aa}. SSA appears in scattering process when one of the colliding proton or the target is transversely polarized with respect to the scattering plane. It can provide information on the three-dimensional structure of the nucleons. From the theoretical point of view, to explain the SSA requires the nonperturbative correlators of quark or gluon and there are two methods for it. The first one is the generalized parton model (GPM) \cite{Ji:2004xq,Ji:2004wu}, where the inclusive cross section is written as a convolution of Transverse Momentum Dependent Partonic Distribution Functions (TMD-PDFs), Transverse Momentum Dependent Fragmentation Functions (TMD-FFs) and QCD partonic cross sections. This method is phenomenologically well studied in refs \cite{Ji:2004xq,Ji:2004wu,GarciaEchevarria:2011rb,Bacchetta:2006tn,Anselmino:2002pd,Boer:1999mm,Arnold:2008kf,Boer:1997mf,Anselmino:2007fs}. The second approach describes the SSA as a twist-3 effect in the collinear factorization, and is suited for describing SSA in the large $\rm p_T$ region. This formalism was originally proposed and further developed by  \cite{Efremov:1984ip,Qiu:1998ia,Kanazawa:2000hz,Kouvaris:2006zy,Eguchi:2006mc,Kanazawa:2014dca}. TMD factorization extends collinear factorization by accounting for the parton transerve momentum, generally denoted by $\rm {\bf k}_T$. It applies to processes in which a momentum transfer is much larger than $\rm {\bf k}_T$, for instance at the LHC when a pair of particles (e.g. two quarkonium states $\rm \mathcal{Q}$) is produced with a large invariant mass ($\rm M_{\mathcal{Q}\mathcal{Q}}$) or a large individual transverse momenta as compared to the sum transverse momenta ($\rm P_{\mathcal{Q}\mathcal{Q}T}$) \cite{Lansberg:2017dzg}.

Among the single spin asymmetries, the Sivers asymmetry plays a vital important role and it is well studied both theoretically and experimentally. The Sivers function \cite{Sivers:1989cc} represents an azimuthal dependence of the number density of unpolarized quarks inside a transversely polarized proton,
and has been measured at HERMES \cite{Airapetian:2004tw,Airapetian:2009ae,Airapetian:2013bim}, COMPASS \cite{Adolph:2012sn,Adolph:2014fjw,Adolph:2017pgv,Aghasyan:2017jop},
JLAB \cite{Qian:2011py,Zhao:2014qvx} and RHIC \cite{Adamczyk:2015gyk} respectively.
It has been found that the initial and final state interactions (gauge links) contribute to the Sivers asymmetries significantly, therefore, they are process dependent \cite{Boer:2003cm}. For example, the Sivers function probed in semi-inclusive deep inelastic scattering (SIDIS) is expected to be the same in magnitude but opposite in sign compared to the one probed in the Drell-Yan (DY) process. Among the Sivers functions, the quark Sivers functions have been widely studied over the years \cite{Anselmino:2016uie}, however, the gluon Sivers function (GSF) still remains poorly measured. An indirect estimation of the GSF was obtained, within the GPM framework in Ref.\cite{DAlesio:2015fwo}, by fitting the midrapidity data on SSA in $\rm \pi^0$ production at RHIC.

Quarkonium processes, through both single and double productions, can be used to probe gluons inside hadron \cite{Brambilla:2010cs}. For example, through single quarkonium production in electron-proton ($\rm ep$) \cite{Godbole:2012bx,Mukherjee:2016qxa,Anselmino:2016fhz,Boer:2016fqd} and proton-proton ($\rm pp$) \cite{Anselmino:2004nk,Godbole:2017syo,DAlesio:2017rzj} collision, the GSFs are comprehensively studied.
The formation of quarkonium out of the two heavy quarks is a nonperturbative process and is treated in terms of different models where the non-relativistic QCD (NRQCD) \cite{Bodwin:1994jh} factorization is one of them.
It has been successfully used to explain the $\rm J/\psi$ hadroproduction at Tevatron \cite{Abe:1997jz,Acosta:2004yw}, 
also data from $\rm J/\psi$ photoproducton at HERA \cite{Aaron:2010gz,Chekanov:2002at,Abramowicz:2012dh}. In the NRQCD, the production and decay of heavy quarkonium are factorized into two stages. First, a heavy quark-antiquark pair is perturbatively built at short distances, which is worked out by expansion in the strong coupling constant $\rm \alpha_s$. Then, the pair nonperturbatively evolves into quarkonium at a long distance. The short distance coefficients are calculated perturbatively by the projection technique 
and the long distance matrix elements (LDMEs) are extracted from the experimental data. The LDMEs behave as powers of v, which is the typical heavy-quark (or antiquark) velocity in the quarkonium rest frame \cite{Lepage:1992tx}. Hence, the NRQCD factorization expression can be thought of as double expansions in terms of $\rm v$ as well as $\rm \alpha_s$. In fact, the asymmetry is very sensitive to the production mechanism. 

Sivers effect has been studied theoretically in several processes, i.e., in $\rm ep^\uparrow$ collision, the heavy quark pair and dijet production \cite{Boer:2016fqd}, the inelastic $\rm J/\psi$ photoproduction \cite{Rajesh:2018qks}, 
$\rm e+p^\uparrow \to e+J/\psi+X$ \cite{Godbole:2012bx,Godbole:2013bca,Godbole:2014tha},
and also in $\rm pp^\uparrow$ scattering, i.e., $\rm pp^\uparrow \to J/\psi+X$ \cite{Godbole:2017syo}, D-meson production \cite{Anselmino:2004nk,Godbole:2016tvq} 
and back to back jet correlations \cite{Boer:2003tx}, etc.
It is found that some processes are not safe in measuring gluon Sivers function due to the problem of TMD factorization breaking contributions \cite{Rogers:2010dm},
though in some aspects they do probe TMDs. On the other hand, among the hadronic collisions the processes 
having one or two color singlets in the final state would, in any cases, be safe \cite{DAlesio:2015fwo} in measuring the GSF.
Therefore, both single and double heavy quarkonium productions are considered to be clean as probes of the GSF.

In this work, we investigate feasibility of using double charmonium production to obtain information on the Sivers function and present predictions for SSA through the process $\rm pp^\uparrow \to J/\psi+J/\psi+X$. Then we estimate the asymmetry by using NRQCD framework with both color singlet and color octet contributions in pp collision. We provide predictions on asymmetry for future proposed experiments at AFTER@LHC which belong to fixed target experiments with $\rm \sqrt{s}=115$ GeV, and also for $\rm \sqrt{s}=200,\ 500$ GeV which will be explored at the RHIC. Two recent extractions \cite{DAlesio:2015fwo,Anselmino:2016uie} were used for the gluon Sivers function from the SSA data in the pp collision at the RHIC. The paper is organized as follows. SSA of double quarkonium production is presented in Sec.\ref{framework}.
In Sec.\ref{numerical} we present the numerical results along with the conclusion in Sec.\ref{summary}.

\section{theoretical framework}
\label{framework}

Single spin asymmetry for the inclusive process $\rm A^\uparrow + B \to C + X$ is described as
\begin{eqnarray}\label{SSA}
\rm A_N = \frac{d\sigma^\uparrow-d\sigma^\downarrow}{d\sigma^\uparrow+d\sigma^\downarrow}
\end{eqnarray}
where $\rm d\sigma^{\uparrow(\downarrow)}$ denotes the differential cross section for scattering of a transversely polarized hadron A off an unpolarized hadron B, with A upwards (downwards) transversely polarized with respect to the scattering plane.

\subsection{Model for $ J/\psi$ pair production}

We consider first the inclusive production of a quarkonium pair in unpolarized proton-proton scattering
\begin{eqnarray}\label{process}
\rm p(P_a)+p(P_b) \to \mathcal{Q}_1(p_1)+\mathcal{Q}_2(p_2)+X
\end{eqnarray}
where $\rm P_i$ and $\rm p_i$ are the four momenta of the particles given in parentheses. We assume that the heavy quark-antiquark pairs produced in the final state are in a bound state described by a nonrelativistic wave function with spin S=1, orbital angular momentum L=0 and total angular momentum J=1. In the following, we adopt the spectroscopic notation $\rm \mathcal{Q}=Q\bar{Q}[^{2S+1} L_J^{(1,8)}]$ with $\rm Q=c$, 
where the color assignments for the quark pair are generally specified by the singlet or octet superscripts, (1) or (8).
The leading contribution must be the color-singlet channel $\rm \mathcal{Q}[^3 S_1^{(1)}]+\mathcal{Q}[^3 S_1^{(1)}]$. 
The color octet channel $\rm \mathcal{Q}[^3 S_1^{(8)}]+\mathcal{Q}[^3 S_1^{(8)}]$ is suppressed by a velocity factor compared to the color singlet channel.
Nevertheless, the color octet channel can also contribute with a large enhancement factor in some kinematic condition \cite{Ko:2010xy}. While the remaining color octet channels suppressed by velocity factors do not have the enhancement factor. Therefore, the contributions to the quarkonium pair production that we consider in this work are $\rm \mathcal{Q}[^3 S_1^{(1)}]+\mathcal{Q}[^3 S_1^{(1)}]$ and $\rm \mathcal{Q}[^3 S_1^{(8)}]+\mathcal{Q}[^3 S_1^{(8)}]$.
This consideration is also suitable for the polarized case.

TMD factorization requires restriction to the so-called correlation limit where the sum transverse momentum of the $\rm J/\psi$'s is small compared to the individual transverse momenta. The individual $\rm p_{T}$ of one of the $\rm J/\psi$ (for example, $\rm p_{1T}$), needs to be large and considered fixed by 5 GeV ($\rm \gg \Lambda_{QCD}$) to avoid evolution as this is related to the hard scale, and the transverse momentum of the other $\rm J/\psi$ ($\rm p_{2T}$) is restricted by $\rm p_{2T} \gg \Lambda_{QCD}$. For the sum transverse momentum of the $\rm J/\psi$'s we have $\rm q_T \sim \Lambda_{QCD}$, where $\rm \Lambda_{QCD}$ is the QCD scale.

In order to guarantee the TMD factorization, the cross section of double $J/\psi$ production should be differential in the sum transverse momentum($\rm P_{\mathcal{Q}\mathcal{Q}T}$)  \cite{Dunnen:2014eta}. Within the generalized parton model formalism, the differential cross section for the process is given by
\begin{eqnarray}\label{unpol} 
\rm  \nonumber
\frac{d\sigma}{d\Omega dy d^2 {\bf q}_T} &=& 
\int \frac{1}{16\hat{s}^2\pi^2} d^2 {\bf k}_{\perp a} d^2 {\bf k}_{\perp b} \delta^2({\bf k}_{\perp a}+{\bf k}_{\perp b}-{\bf q}_{T}) \frac{p_1^2 dp_1 \delta(p_1 \sin\theta_1-5) }{\sqrt{m_1^2+p_1^2}} dM^2 \delta(q^2-2q \cdot p_1) 
\\ \nonumber
&& ~~~ \bigg[f_{g/p} (x_a,{\bf k_{\perp a}}) f_{g/p}(x_b,{\bf k_{\perp b}}) \Big( \Big|\mathcal{\overline{M}}_{gg\to \mathcal{Q}\mathcal{Q} }[^3 S_1^{(1)},^3 S_1^{(1)}]\Big|^2+\Big|\mathcal{\overline{M}}_{gg\to \mathcal{Q}\mathcal{Q} }[^3 S_1^{(8)},^3 S_1^{(8)}]\Big|^2 \Big)
\\
&+& \sum_{q,\bar{q}} f_{q/p}(x_a,{\bf k_{\perp a}})f_{\bar{q}/p}(x_b,{\bf k_{\perp b}}) \Big( \Big|\mathcal{\overline{M}}_{q\bar{q}\to \mathcal{Q}\mathcal{Q} }[^3 S_1^{(1)},^3 S_1^{(1)}]\Big|^2+\Big|\mathcal{\overline{M}}_{q\bar{q}\to \mathcal{Q}\mathcal{Q} }[^3 S_1^{(8)},^3 S_1^{(8)}]\Big|^2 \Big) \bigg] ,
\end{eqnarray}
where $\rm {\bf k}_{\perp a}$ and $\rm {\bf k}_{\perp b}$ are the transverse momentum of two initial gluons, q and y are the transverse momentum and rapidity of the final system respectively. We have written the four momentum conservation delta function by 
\begin{eqnarray}\label{deltafunc} 
\rm \delta^4 (k_a+k_b-p_1-p_2) = \frac{2}{s}\delta(x_a-\frac{Me^y}{\sqrt{s}})\delta(x_b-\frac{Me^{-y}}{\sqrt{s}})\delta^2 ({\bf k}_{\perp a}+{\bf k}_{\perp b}-{\bf q}_{T}).
\end{eqnarray}
The fractions $\rm x_a$ and $\rm x_b$ of the initial protons' longitudinal momenta are given by
\begin{eqnarray}\label{fractions} 
\rm
x_a=\frac{Me^y}{\sqrt{s}} \qquad \qquad x_b=\frac{Me^{-y}}{\sqrt{s}}
\end{eqnarray}
where M is the invariant mass of the final mesons.
In addition, we define the four-momentum and mass of $\rm \mathcal{Q}_1$ as $\rm p_1$ and $\rm m_1$, the corresponding solid angle is denoted by $\Omega$. The unpolarized gluon TMD, $\rm f_{g/p}$, represents the density of gluons inside an unpolarized proton. $\rm \mathcal{\overline{M}}_{g+g\rightarrow J/\psi + J/\psi}[^3 S_1^{(1)},^3 S_1^{(1)}]$ ($\rm \mathcal{\overline{M}}_{g+g\rightarrow J/\psi + J/\psi}[^3 S_1^{(8)},^3 S_1^{(8)}]$) is the amplitude of gluon-gluon fusion process with two final color singlet(octet) and its square is given in Appendix.\ref{ap1}.
The delta function $\rm \delta(q^2-2q \cdot p_1) $ is solved in Sec.\ref{kin}. In order to focus on the contribution from initial gluons we impose a cut: $\rm s x_a x_b=\hat{s} \gg \Lambda_{QCD}$, since the gluon distribution function is much larger than the quark distribution function in the case with relative large x \cite{Qiu:2011ai}. As a result, the parton process with $\rm q\bar{q}$ initial states is neglected and we only consider gg initial state in the present paper.

\subsection{Kinematics}\label{kin}

We consider a frame in which the unpolarized and polarized protons are moving along $\rm z$ and -$\rm z$-axes respectively. The four momenta of the protons are given by
\begin{eqnarray}\label{beam-mom}
\rm P_a=\frac{\sqrt{s}}{2}(1,0,0,1),\ \ P_b=\frac{\sqrt{s}}{2}(1,0,0,-1).
\end{eqnarray}
The center of mass (CM) energy of proton-proton system is $\rm \hat{s}=(P_a+P_b)^2=M^2$. The above four momenta in light-cone coordinate system can be written as 
\begin{eqnarray}\label{beam-lc}
\rm P_a^\mu=\sqrt{\frac s2}n_+^\mu,~~ P_b^\mu=\sqrt{\frac s2}n_-^\mu,
\end{eqnarray}
where $\rm n_+$ and $\rm n_-$ are two light-like vectors with $\rm n_+.n_-=1$ and $\rm n_+^2=n^2_-=0$, and given by
\begin{eqnarray}\label{lc-vec}
\rm n_+^\mu=\frac{1}{\sqrt{2}}(1,0,0,1),~~~n_-^\mu=\frac{1}{\sqrt{2}}(1,0,0,-1),
\end{eqnarray}
where $\rm x_a=\frac{k_a^+}{P_a^+}$ is the light-cone momentum fraction.  The four momentum of one $\rm \rm J/\psi$ is given by
\begin{eqnarray}\label{ini-mom}
\rm 
p_1^\mu=\left( \sqrt{m_1^2+p_1^2}, p_1\sin\theta_1 \cos\phi_1, p_1\sin\theta_1 \sin\phi_1, p_1 \cos\theta_1 \right)
\end{eqnarray}
and the sum 4-momentum of double $\rm J/\psi$ is parameterized as
\begin{eqnarray}\label{sum-mom}
\rm 
q^\mu = \left( \sqrt{M^2+q_{T}^2} \cosh y, q_{T}\cos\phi, q_{T}\sin\phi, \sqrt{M^2+q_{T}^2} \sinh y \right).
\end{eqnarray}
Then the gluons four momenta are given by
\begin{eqnarray} \nonumber\label{par-mom}
\rm k_a^\mu&=&\rm 
x_a \sqrt{\frac s2}n_+^\mu + \frac{k^2_{\perp a}}{2x_2 \sqrt{\frac s2}}n_-^\mu + {\bf k}^\mu_{\perp a} =\left( \frac{x_a \sqrt{s}}{2}+\frac{k_{\perp a}^2}{2x_a \sqrt{s}}, k_{\perp a}\cos\phi_a, k_{\perp a}\sin\phi_a, \frac{x_a \sqrt{s}}{2}-\frac{k_{\perp a}^2}{2x_a \sqrt{s}}  \right) 
\\
\rm k_{\perp b}^\mu&=&\rm 
(0,k_{\perp b}\cos\phi_b,k_{\perp b}\sin\phi_b,0)=(0,q_T \cos\phi-k_{\perp a}\cos\phi_a,q_T\sin\phi-k_{\perp a}\sin\phi_a,0).
\end{eqnarray}
By using the above relations, we can work out the delta function $\rm \delta (q^2-2q \cdot p_1)$ as
\begin{eqnarray}\label{deltafunc2} \nonumber
&&
\rm \int \delta(q^2-2q p_1) f(M^2) dM^2
\\ \nonumber
&=&\rm 
\int dM^2 \delta\bigg(M^2-2\left[ \sqrt{M^2+q_{T}^2} \Big(\cosh y\sqrt{m_1^2+p_1^2} - \sinh y \, p_1 \cos\theta_1 \Big)-q_T p_1 \sin\theta_1 \cos(\phi-\phi_1) \right]\bigg)
\\
&=&\rm 
\frac{2t_0 f(t_0^2-q_T^2)}{\Big|2t_0-2\Big( \cosh y\sqrt{m_1^2+p_1^2} - \sinh y \, p_1 \cos\theta_1 \Big)\Big|}
\end{eqnarray}
where $\rm t_0$ satisfies
\begin{eqnarray} \label{quad-eq} 
\rm 
\bigg(t_0^2-q_T^2-2\left[ t_0 \Big(\cosh y\sqrt{m_1^2+p_1^2} - \sinh y \, p_1 \cos\theta_1 \Big)-q_T p_1 \sin\theta_1 \cos(\phi-\phi_1) \right]\bigg) = 0
\end{eqnarray}
and $\rm f(M^2)$ is a function with respect to $\rm M^2$. Then we can write down the expressions of Mandelstam variables as below
\begin{eqnarray}\label{man-var} \nonumber
\rm \hat{s}& =&\rm t_0^2-q_T^2 \\ \nonumber
\rm \hat{t}& =&\rm (k_1-p_1)^2 \\
\rm \hat{u}& =&\rm 2m_1^2-\hat{s}-\hat{t} .
\end{eqnarray}

\subsection{The Sivers effect}

Now, we are in the position to write down the expression of numerator and denominator terms of Eq.(\ref{SSA}) when the target proton is polarized. The expression reads
\begin{eqnarray}\label{num-ssa} \nonumber
\rm 
d\sigma^{\uparrow}-d\sigma^{\downarrow}&=&\rm
\frac{d\sigma^{pp^{\uparrow}\rightarrow J/\psi J/\psi X}}{d\Omega dy d^2 {\bf q}_T}-\frac{d\sigma^{pp^{\downarrow}\rightarrow J/\psi J/\psi X}}{d\Omega dy d^2 {\bf q_T}}  \\ \nonumber
&=& \rm  
\int \frac{1}{16\hat{s}^2\pi^2} d^2 {\bf k}_{\perp a} d^2 {\bf k}_{\perp b} \delta^2({\bf k}_{\perp a}+{\bf k}_{\perp b}-{\bf q}_{T}) \frac{p_1^2 dp_1 \delta(p_1 \sin\theta_1-5) }{\sqrt{m_1^2+p_1^2}} dM^2 \delta(q^2-2q \cdot p_1)  \\ 
&&\rm f_{g/p} (x_a,{\bf k_{\perp a}}) \Delta^N f_{g/p^{\uparrow}}(x_b,{\bf k_{\perp b}})  \Big|\mathcal{\overline{M}}\Big|^2
\end{eqnarray}
and 
\begin{eqnarray}\label{den-ssa} \nonumber
\rm 
d\sigma^{\uparrow}+d\sigma^{\downarrow}&=&\rm
\frac{d\sigma^{pp^{\uparrow}\rightarrow J/\psi J/\psi X}}{d\Omega dy d^2 {\bf q}_T}+\frac{d\sigma^{pp^{\downarrow}\rightarrow J/\psi J/\psi X}}{d\Omega dy d^2 {\bf q_T}} \\ \nonumber
&=&\rm 
2\int \frac{1}{16\hat{s}^2\pi^2} d^2 {\bf k}_{\perp a} d^2 {\bf k}_{\perp b} \delta^2({\bf k}_{\perp a}+{\bf k}_{\perp b}-{\bf q}_{T}) \frac{p_1^2 dp_1 \delta(p_1 \sin\theta_1-5) }{\sqrt{m_1^2+p_1^2}} dM^2 \delta(q^2-2q \cdot p_1) 
\\ 
&& f_{g/p} (x_a,{\bf k_{\perp a}}) f_{g/p}(x_b,{\bf k_{\perp b}})  
\Big|\mathcal{\overline{M}}\Big|^2 
\end{eqnarray} 
where $\rm \Delta^N f_{g/p^{\uparrow}}(x_g,{\bm k}_{\perp g})$, GSF, represents the parton density function of gluon inside the transversely polarized proton and is defined as below
\begin{equation}\label{gsf}
\rm 
\Delta^Nf_{g/p^{\uparrow}}(x_b,{\bf k}_{\perp b})
= f_{g/p^{\uparrow}}(x_b,{\bf k}_{\perp b})- f_{g/p^{\downarrow}}(x_b,{\bf k}_{\perp b})=\Delta^Nf_{g/p^{\uparrow}}(x_b,k_{\perp b})~{\hat{\bf S}}\cdot(\hat{\bf P}\times\hat{\bf 
	k}_{\perp b}).
\end{equation}
In estimating the SSA numerically, we have to discuss about the parameterization of TMDs.
In this paper, we assume that the unpolarized gluon TMD follows the general Gaussian distribution. The Gaussian parameterization of unpolarized TMD is
\begin{equation}\label{gaus}
\rm 
f_{g/p}(x_a,{\bf k}_{\perp a})=f_{g/p}(x_a)\frac{1}{\pi \langle k^2_{\perp a}\rangle}
e^{-{\bf k}^2_{\perp a}/\langle k^2_{\perp a}\rangle}.
\end{equation}
Here, $\rm x_a$ and $\rm k_{\perp a}$ dependencies of the TMD are factorized. The collinear PDF is denoted by $\rm  f_{g/p}(x_a)$ which is measured at the fixed scale $\rm  \mu=\sqrt{m_1^2+p_{1T}^2}$. 

We choose a frame as discussed in the last subsection where the polarized proton is moving along -z axis with momentum $\rm {\bf P}$, and transversely polarized $\rm \hat{\bf S}=(\cos\phi_s, \sin\phi_s,0)$. The transverse momentum of the initial polarized gluon is $\rm {\bf k}_{\perp b}=k_{\perp b}(\cos\phi_b,\sin\phi_b,0)$,
then
\begin{equation}\label{cro-pro}
\rm 
\hat{{\bf S}}\cdot(\hat{\bf P}\times\hat{\bf k}_{\perp g})=\sin(\phi_b-\phi_s).
\end{equation}
We have taken $\rm \phi_s=\pi/2$ for numerical estimation.
The parameterization of GSF is given by \cite{DAlesio:2015fwo,Anselmino:2016uie}
\begin{equation}\label{gsf2}
\rm 
\Delta^{N}f_{g/p^{\uparrow}}(x_b, k_{\perp b})=2\mathcal{N}_g(x_b)f_{g/p}(x_b)h(k_{\perp b})
\frac{e^{-{\bf k}^2_{\perp b}/\langle k^2_{\perp b}\rangle}}{\pi\langle k^2_{\perp b}\rangle},
\end{equation}
here $\rm f_{g/p}(x_b)$ is the usual collinear gluon PDF and 
\begin{equation}\label{ngx}
\rm 
\mathcal{N}_g(x_b)=N_g x_b^\alpha(1-x_b)^\beta\frac{(\alpha+\beta)^{(\alpha+\beta)}}{\alpha^\alpha\beta^\beta}.
\end{equation}
The definition of function $\rm h(k_{\perp b})$ is given by
\begin{equation}\label{h}
\rm 
h(k_{\perp b})=\sqrt{2e}\frac{k_{\perp b}}{M_1}e^{-{\bf k}^2_{\perp b}/M^2_1}.
\end{equation}
The $\rm k_{\perp b}$ dependent part of Sivers function can be rewritten as
\begin{equation}
\rm h(k_{\perp b})\frac{e^{-{\bf k}^2_{\perp b}/\langle k^2_{\perp b}\rangle}}{\pi\langle k^2_{\perp b}\rangle}
=\frac{\sqrt{2e}}{\pi}\sqrt{\frac{1-\rho}{\rho}}k_{\perp b}
\frac{e^{-{\bf k}^2_{\perp b}/\rho\langle k^2_{\perp b}\rangle}}{\langle k^2_{\perp b}\rangle^{3/2}},
\end{equation}
where we defined 
\begin{equation}\label{rho}
\rm \rho=\frac{M^2_1}{\langle k^2_{\perp b}\rangle + M^2_1}.
\end{equation}
At RHIC, the GSF from pion production data has been extracted by D'Alesio et al. \cite{DAlesio:2015fwo}, and we define two sets of two best fit parameters obtained as SIDIS1 and SIDIS2.
In addition, the quark and anti-quark Sivers functions have been extracted by Anselmino et al. \cite{Anselmino:2016uie} using the latest SIDIS data. However, GSF has not been extracted yet from SIDIS data. Therefore, in order to estimate the asymmetry, we have to use some parameterizations \cite{Boer:1999mm} listed below to reach the best fit parameters of gluon Sivers function \begin{eqnarray} \label{bva-bvb}
\rm (a)~~\mathcal{N}_g(x_b)&=&\rm \left(\mathcal{N}_u(x_b)+\mathcal{N}_d(x_b)\right)/2 \nonumber\\
\rm (b)~~\mathcal{N}_g(x_b)&=&\rm \mathcal{N}_d(x_b).
\end{eqnarray}
We call the parameterization (a) and (b) as BV-a and BV-b respectively. The best fit parameters are 
tabulated in \tablename{ \ref{table1}}.
\begin{table}[htb!]
	\setlength{\tabcolsep}{5mm}
	\begin{center}
		\begin{tabular}{|c|c|c|c|c|c|c|c|}
			\hline
			  & $\rm N_a$ & $\alpha$ & $\beta$ & $\rm M_1^2$ GeV$^2$ & $\rho$ & $\rm \langle k_\perp^2 \rangle$\ GeV$^2$ & Notation \\
			\hline
			g & 0.65 & 2.8 & 2.8 &  & 0.687 & 0.25 & SIDIS1 \\
			\hline
			g & 0.05 & 0.8 & 1.3 &  & 0.576 & 0.25 & SIDIS2 \\
			\hline
			u & 0.18 & 1.0 & 6.6 & 0.8 &  & 0.57 & BV-a \\
			\hline
			d & -0.52 & 1.9 & 10.0 & 0.8 &  & 0.57 & BV-b \\
			\hline
		\end{tabular}
	\end{center}
	\caption{\label{table1}Best fit parameters of Sivers function.}
\end{table}

Generally, we take $\rm \langle k_{\perp a}^2 \rangle = \langle k_{\perp b}^2 \rangle = \langle k_{\perp}^2 \rangle$, then the final expressions of  numerator and denominator terms of Eq.\eqref{SSA} are given by
\begin{eqnarray}\label{final-num} \nonumber
\rm d\sigma^{\uparrow}-d\sigma^{\downarrow}
&=&\rm \int \frac{1}{16\hat{s}^2\pi^2} d^2 {\bf k}_{\perp a} d^2 {\bf k}_{\perp b} \delta^2({\bf k}_{\perp a}+{\bf k}_{\perp b}-{\bf q}_{T}) \frac{p_1^2 dp_1 \delta(p_1 \sin\theta_1-5) }{\sqrt{m_1^2+p_1^2}} dM^2 \delta(q^2-2q \cdot p_1) 
\\
&&\rm f_{g/p}(x_a) \frac{1}{\pi \langle k^2_{\perp }\rangle}
e^{-{\bf k}^2_{\perp a}/\langle k^2_{\perp}\rangle} 2\mathcal{N}_g(x_b)f_{g/p}(x_b) \frac{\sqrt{2e}}{\pi}\sqrt{\frac{1-\rho}{\rho}}k_{\perp b} 
\frac{e^{-{\bf k}^2_{\perp b}/\rho\langle k^2_{\perp}\rangle}}{\langle k^2_{\perp}\rangle^{3/2}}  \Big|\mathcal{\overline{M}}\Big|^2 \sin(\phi-\phi_s)
\end{eqnarray}
and
\begin{eqnarray}\label{final-den} \nonumber
\rm d\sigma^{\uparrow}+d\sigma^{\downarrow}
&=&\rm 2\int \frac{1}{16\hat{s}^2\pi^2} d^2 {\bf k}_{\perp a} d^2 {\bf k}_{\perp b} \delta^2({\bf k}_{\perp a}+{\bf k}_{\perp b}-{\bf q}_{T}) \frac{p_1^2 dp_1 \delta(p_1 \sin\theta_1-5) }{\sqrt{m_1^2+p_1^2}} dM^2 \delta(q^2-2q \cdot p_1) 
\\
&&\rm f_{g/p}(x_a) \frac{1}{\pi \langle k^2_{\perp }\rangle}
e^{-{\bf k}^2_{\perp a}/\langle k^2_{\perp}\rangle} 
f_{g/p}(x_b) \frac{1}{\pi \langle k^2_{\perp }\rangle}
e^{-{\bf k}^2_{\perp b}/\langle k^2_{\perp}\rangle}
  \Big|\mathcal{\overline{M}}\Big|^2 .
\end{eqnarray}
Then we reach the Sivers asymmetry 
\begin{eqnarray}\label{final-ssa}
\rm 
A_N^{\sin(\phi-\phi_s)}=2\frac{\int d\phi [d\sigma^\uparrow-d\sigma^\downarrow]\sin(\phi-\phi_s)}{\int d\phi [d\sigma^\uparrow+d\sigma^\downarrow]}
\end{eqnarray}
where $\phi$ is the azimuthal angle of double $\rm J/\psi$ system.

\section{Numerical results}
\label{numerical}

In this section, we present predictions of transverse single spin asymmetry in $\rm p+p^\uparrow \to J/\psi+J/\psi+X$, obtained using the recent direct fits \cite{DAlesio:2015fwo} and the BV models \cite{Boer:2003tx} of the GSF with the corresponding best fit parameters of quark Sivers functions.
The GSF parameterizations denoted by "SIDIS1" and "SIDIS2" are given in refs \cite{Anselmino:2005ea,Anselmino:2008sga}. MSTW2008 \cite{Martin:2009iq} is used for collinear PDFs probed at the fixed scale $\rm \mu = \sqrt{m_1^2+p_{1T}^2}$, and $\rm m_1 = 3.096$ GeV is the mass of $\rm J/\psi$. We use the wave function at origin and the LDME $\rm \langle O_8(^3 S_1) \rangle$ whose values are listed as \cite{Li:2009ug,Ko:2010xy}
\begin{equation}\label{lme}
\begin{aligned}
&\rm |R(0)[^3 S_1]|^2 = 0.815\ \text{GeV}^3
\\
&\rm \langle O_8(^3 S_1) \rangle = 3.9 \times 10^{-3}\ \text{GeV}^3 .
\end{aligned}
\end{equation}
The transverse momentum of the two initial gluons $\rm k_{\perp 1}$, $\rm k_{\perp 2}$ in Eq.(\ref{unpol}) are integrated within the limits $\rm 0<k_{\perp 1}, k_{\perp 2}<1$ GeV. We have noticed that the higher values of $\rm k_{\perp 1 max}$ and $\rm k_{\perp 2 max}$ do not make a difference with the SSA and the unpolarized differential cross section.

Our predictions of SSA are given for three different CM energies $\rm \sqrt{s}$ = 115 GeV (AFTER@LHC), 
$\rm \sqrt{s}$ = 200 GeV (RHIC1) and $\rm \sqrt{s}$ = 500 GeV (RHIC2). As mentioned above, we impose the following overall cuts for three different CM energies so as to ignore $\rm q\bar{q}$ channel contributions to $\rm J/\psi$ pair, and apply TMD factorization:
\begin{equation}\label{ove-cut}
\begin{aligned}
&\rm \hat{s}=M^2 \gg \Lambda_{QCD} \qquad \qquad 0 < q_T < 1\ \text{GeV} \qquad \qquad 4 < p_{2T} < 6\ \text{GeV}.
\end{aligned}
\end{equation}
Moreover, we impose cuts on both individual $\rm J/\psi$'s rapidity $\rm -2.8 < y_1,y_2 < 0.2$ for $\sqrt{s}=115$ GeV, $\rm 2< y_1,y_2 < 3$ and $\rm 3 < y_1,y_2 < 3.8$ for $\rm \sqrt{s}=200$ GeV, and $\rm 3 < y_1,y_2 < 4$ for $\rm \sqrt{s} = 500$ GeV. The given rapidity ranges were chosen keeping in mind the proposed forward sPHENIX (fsPHENIX) upgrade \cite{Barish:2012ha,Aschenauer:2015eha}.
In practice, we abandon the cut $\rm 3 < y_1,y_2 < 3.8$ for $\sqrt{s}=200$ GeV, since it cuts off almost all events.
We present the asymmetry predictions as a function of the sum transverse momentum ($\rm q_T$) and the rapidity ($\rm y$) of two quarkonium system as well as the rapidity difference of two $\rm J/\psi$ ($\rm \Delta y$).

\begin{figure}[hbtp]
\includegraphics[height=6.3cm,width=5.9cm,angle=0]{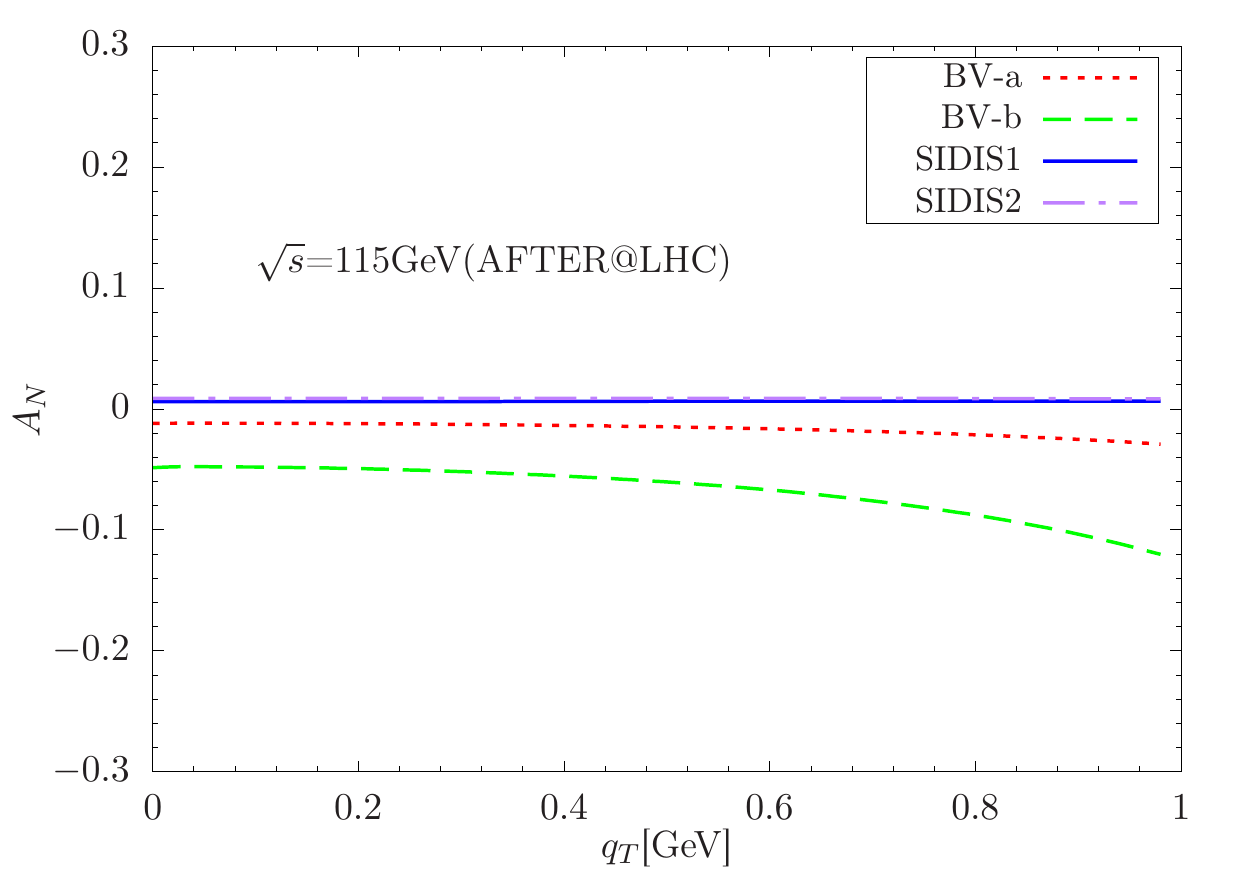}
\includegraphics[height=6.3cm,width=5.9cm,angle=0]{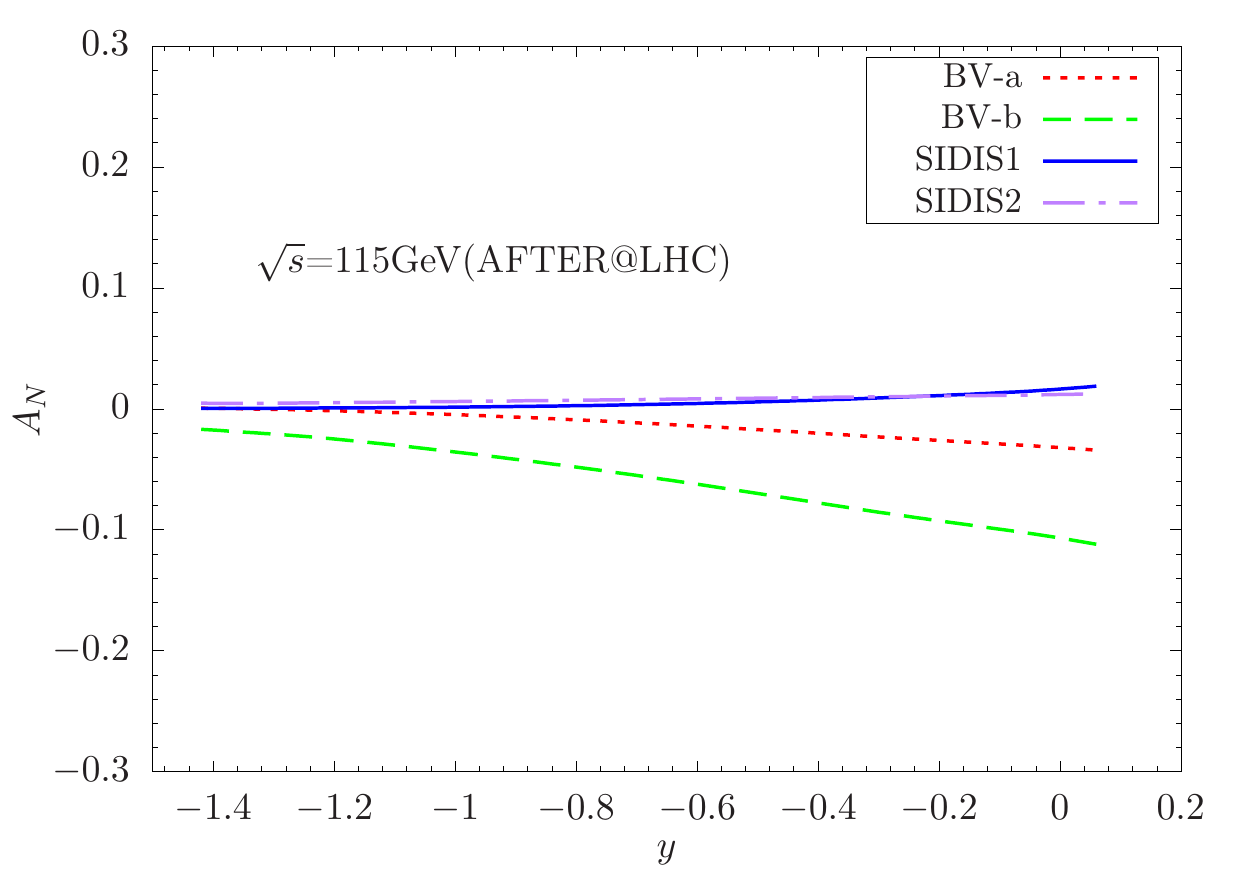}
\includegraphics[height=6.3cm,width=5.9cm,angle=0]{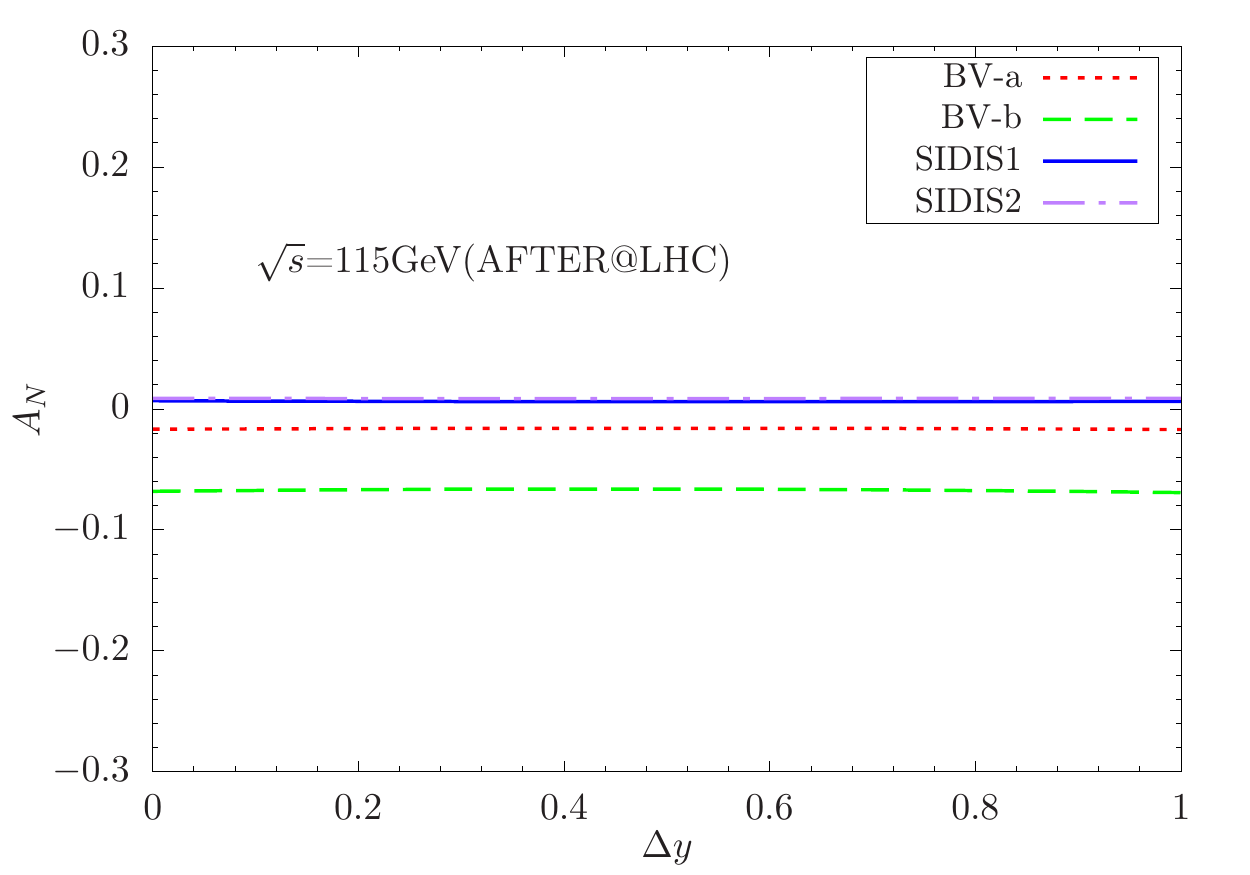}
\caption{\label{115}
Single spin asymmetry in $\rm p+p^\uparrow \to J/\psi+J/\psi+X$ process as functions of $\rm q_T$, y, $\rm \Delta y$ with BV-a, BV-b, SIDIS1 and SIDIS2 parameters at $\rm \sqrt{s}=115$ GeV (AFTER@LHC). The integration ranges are $\rm 0 < q_T < 1$ GeV, and we impose cuts on both individual $\rm J/\psi$'s rapidity $\rm -2.8 < y_1,y_2 < 0.2$.}
\end{figure}

\begin{figure}[htp]
\includegraphics[height=6.3cm,width=5.9cm,angle=0]{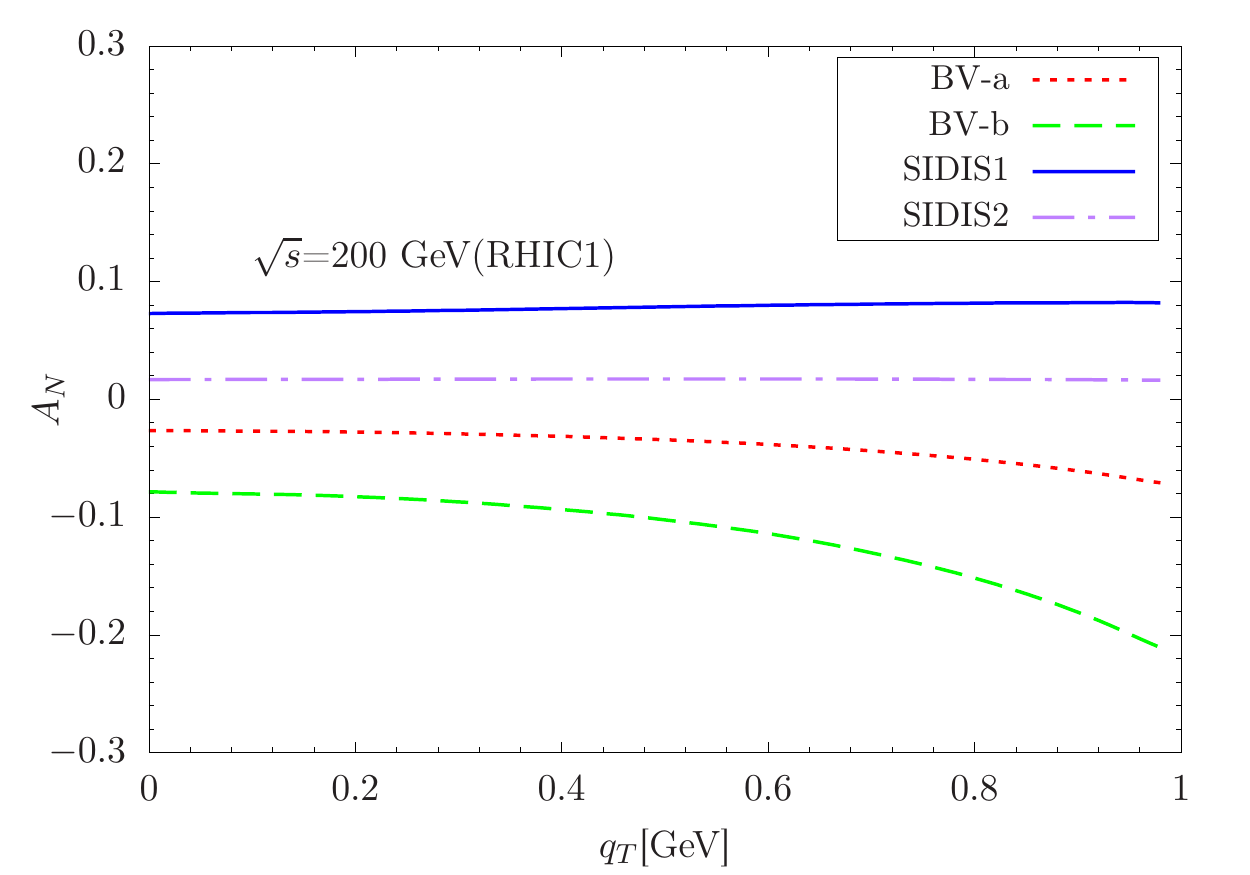}
\includegraphics[height=6.3cm,width=5.9cm,angle=0]{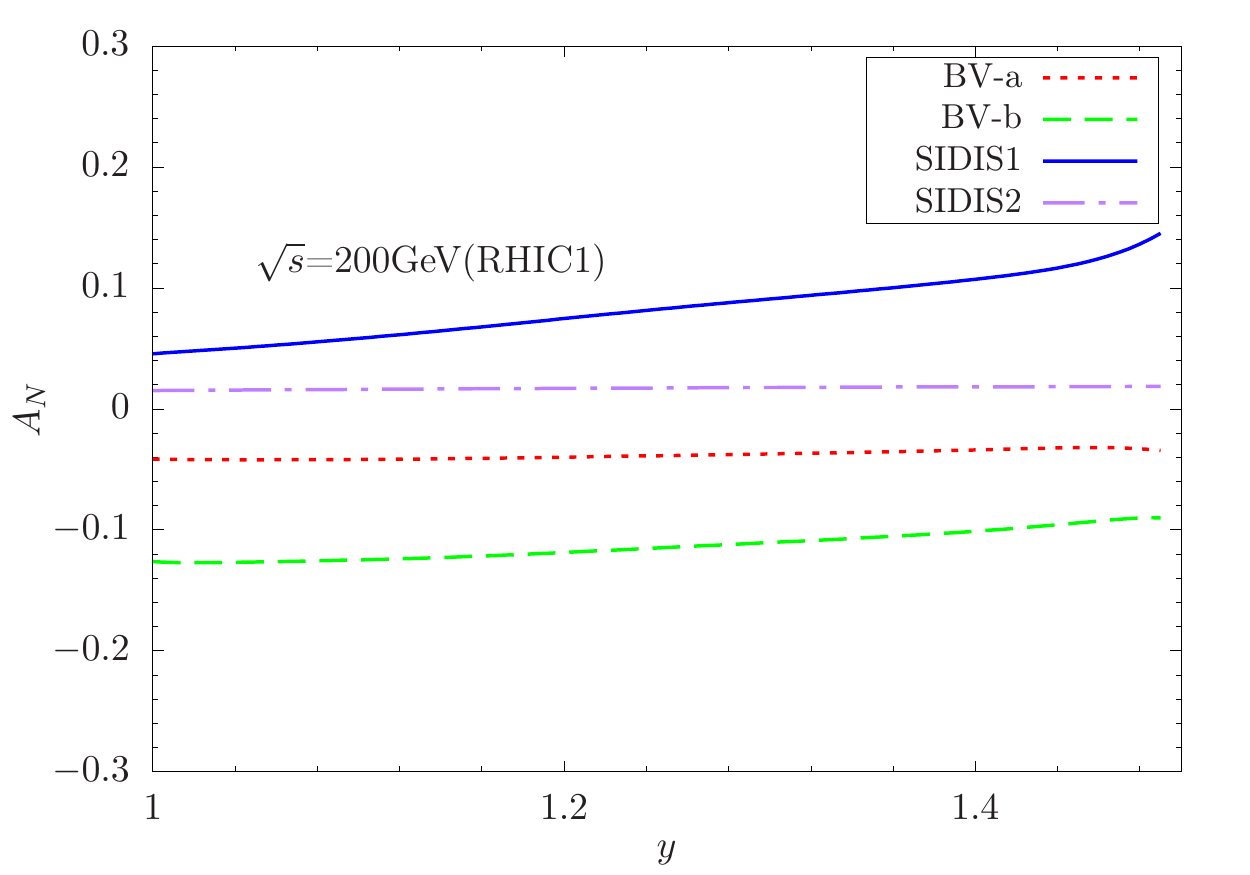}
\includegraphics[height=6.3cm,width=5.9cm,angle=0]{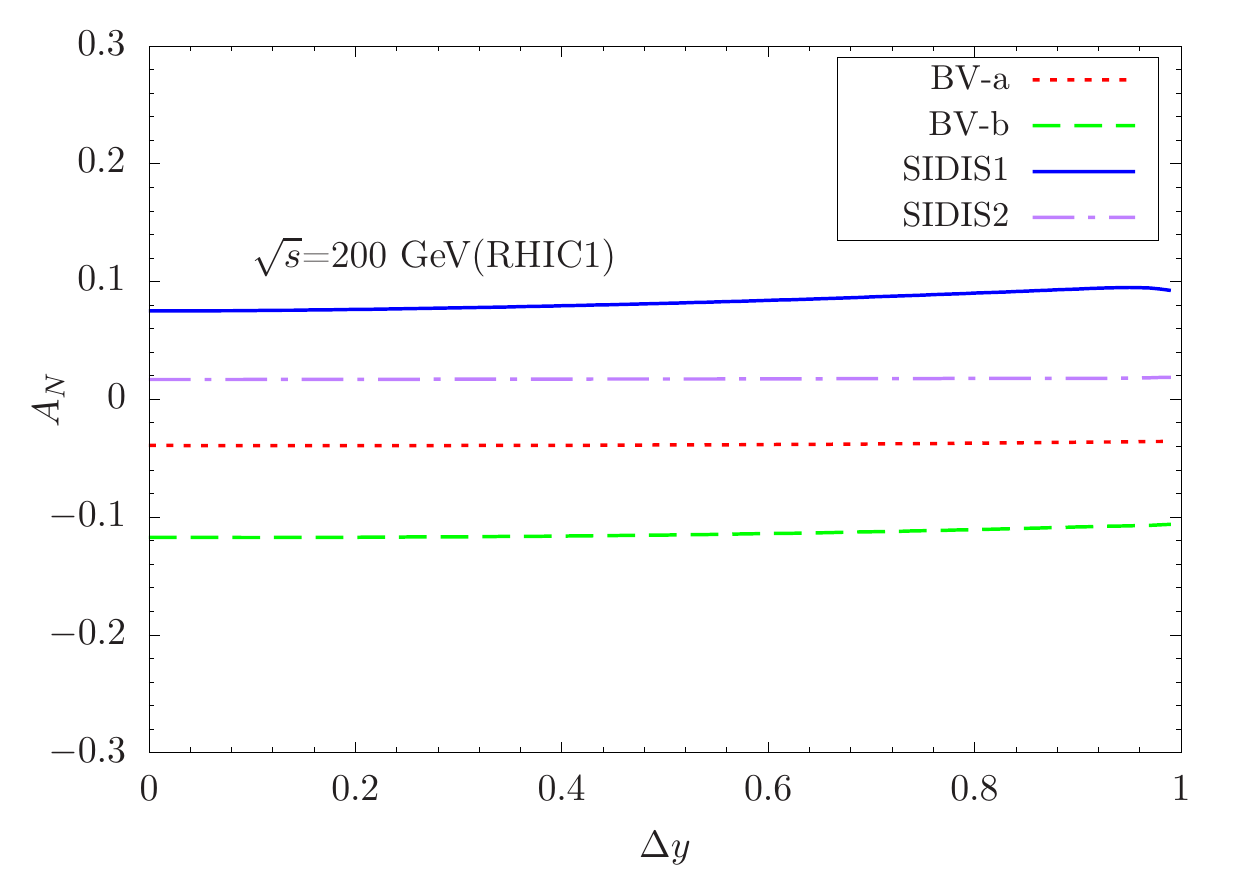}
\caption{\label{200_1}
Single spin asymmetry in $\rm p+p^\uparrow \to J/\psi+J/\psi+X$ process as functions of $\rm q_T$, y, $\rm \Delta y$ with BV-a, BV-b, SIDIS1 and SIDIS2 parameters at $\rm \sqrt{s}=200$ GeV (RHIC1). The integration ranges are $\rm 0 < q_T < 1$ GeV, and we impose cuts on both individual $\rm J/\psi$'s rapidity $\rm 2< y_1,y_2 < 3$.}
\end{figure}

\begin{figure}[hbtp]
\includegraphics[height=6.3cm,width=5.9cm,angle=0]{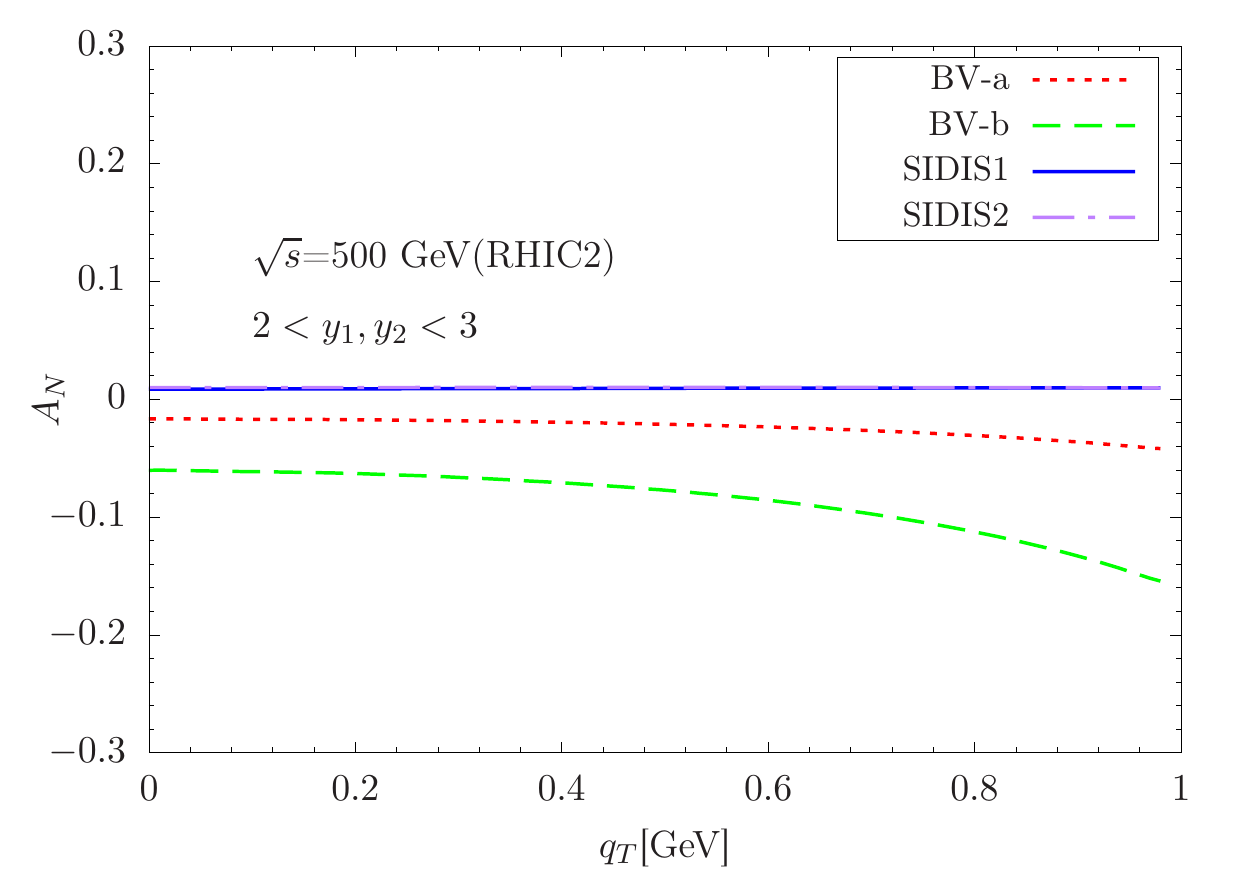}
\includegraphics[height=6.3cm,width=5.9cm,angle=0]{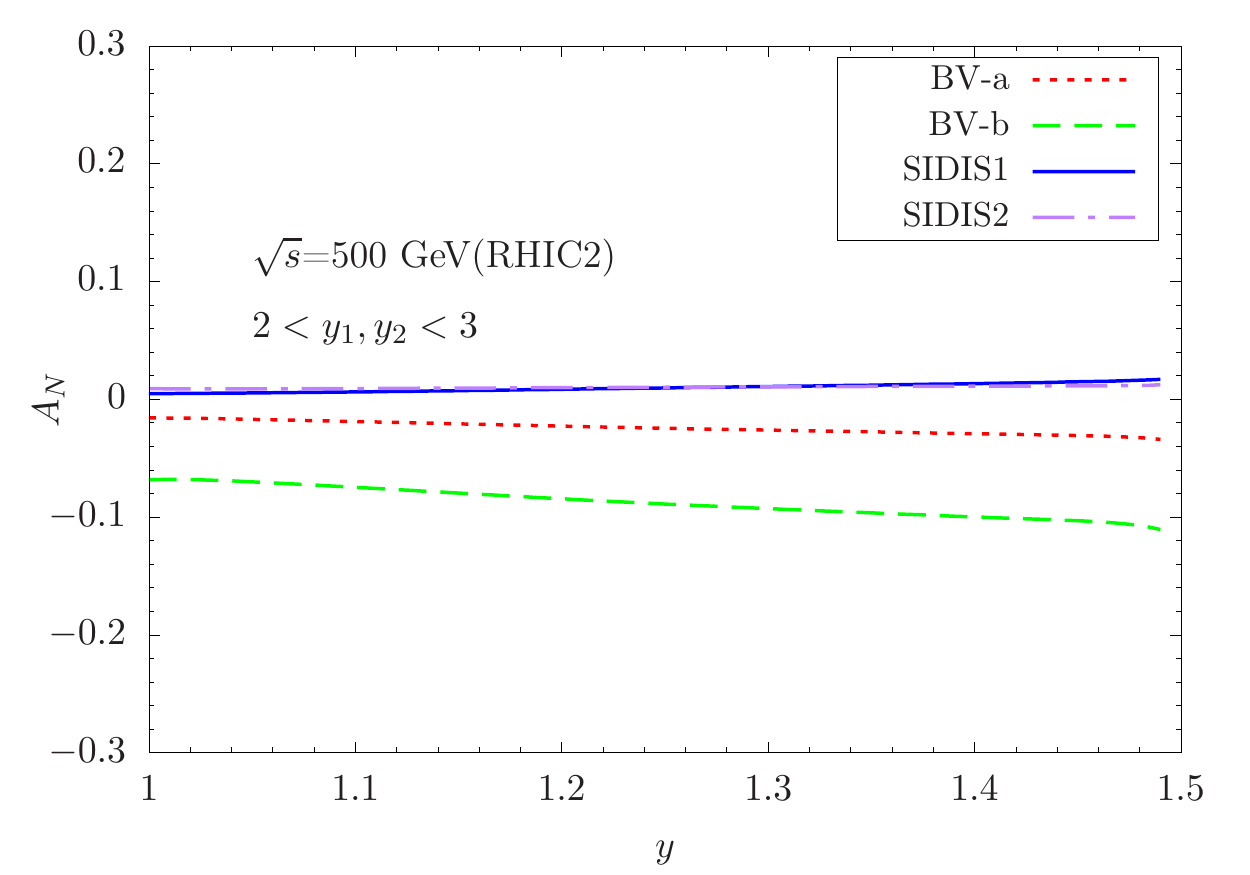}
\includegraphics[height=6.3cm,width=5.9cm,angle=0]{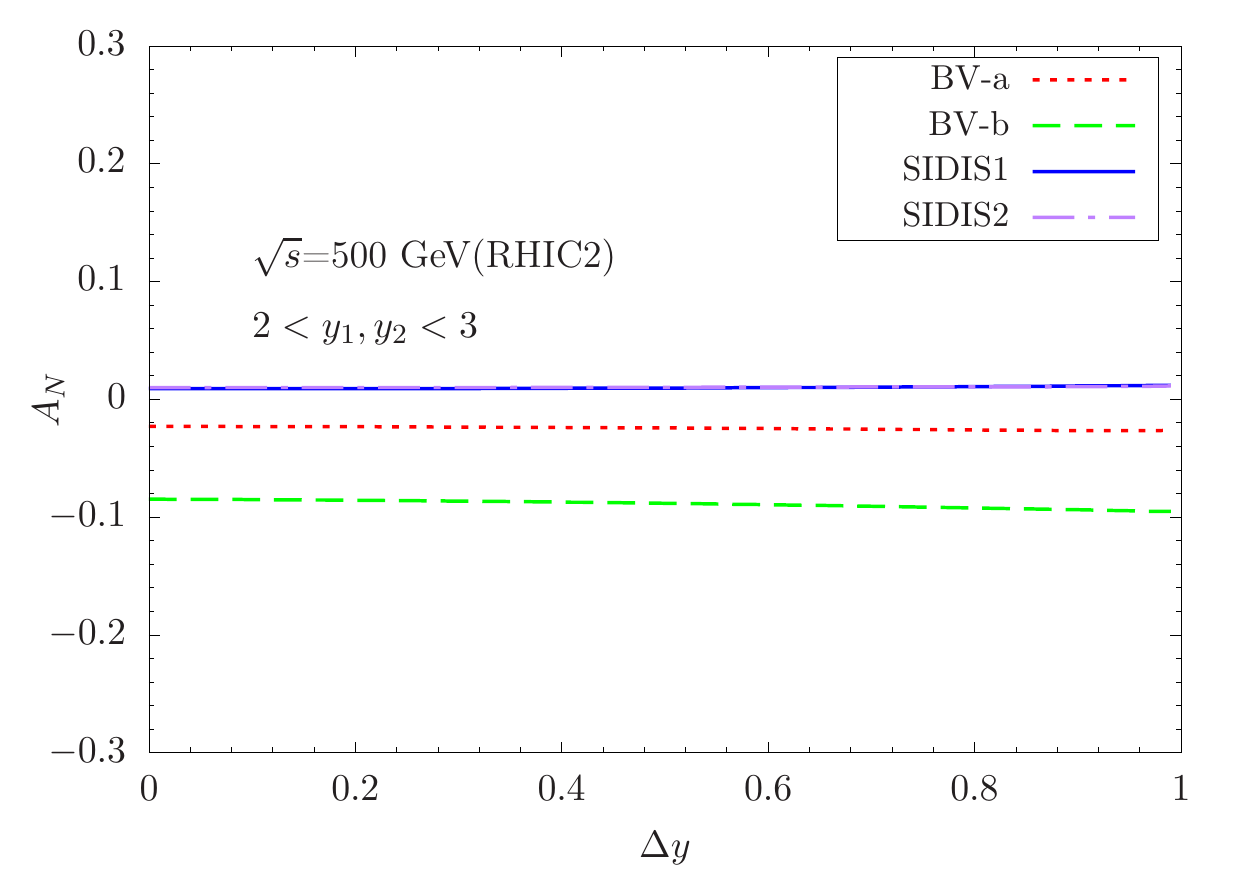}
\caption{\label{500_1}
Single spin asymmetry in $\rm p+p^\uparrow \to J/\psi+J/\psi+X$ process as functions of $\rm q_T$, y, $\rm \Delta y$ with BV-a, BV-b, SIDIS1 and SIDIS2 parameters at $\rm \sqrt{s}=500$ GeV (RHIC2). The integration ranges are $\rm 0 < q_T < 1$ GeV, and we impose cuts on both individual $\rm J/\psi$'s rapidity $\rm 2 < y_1,y_2 < 3$.}
\end{figure}

\begin{figure}[hbtp]
\includegraphics[height=6.3cm,width=5.9cm,angle=0]{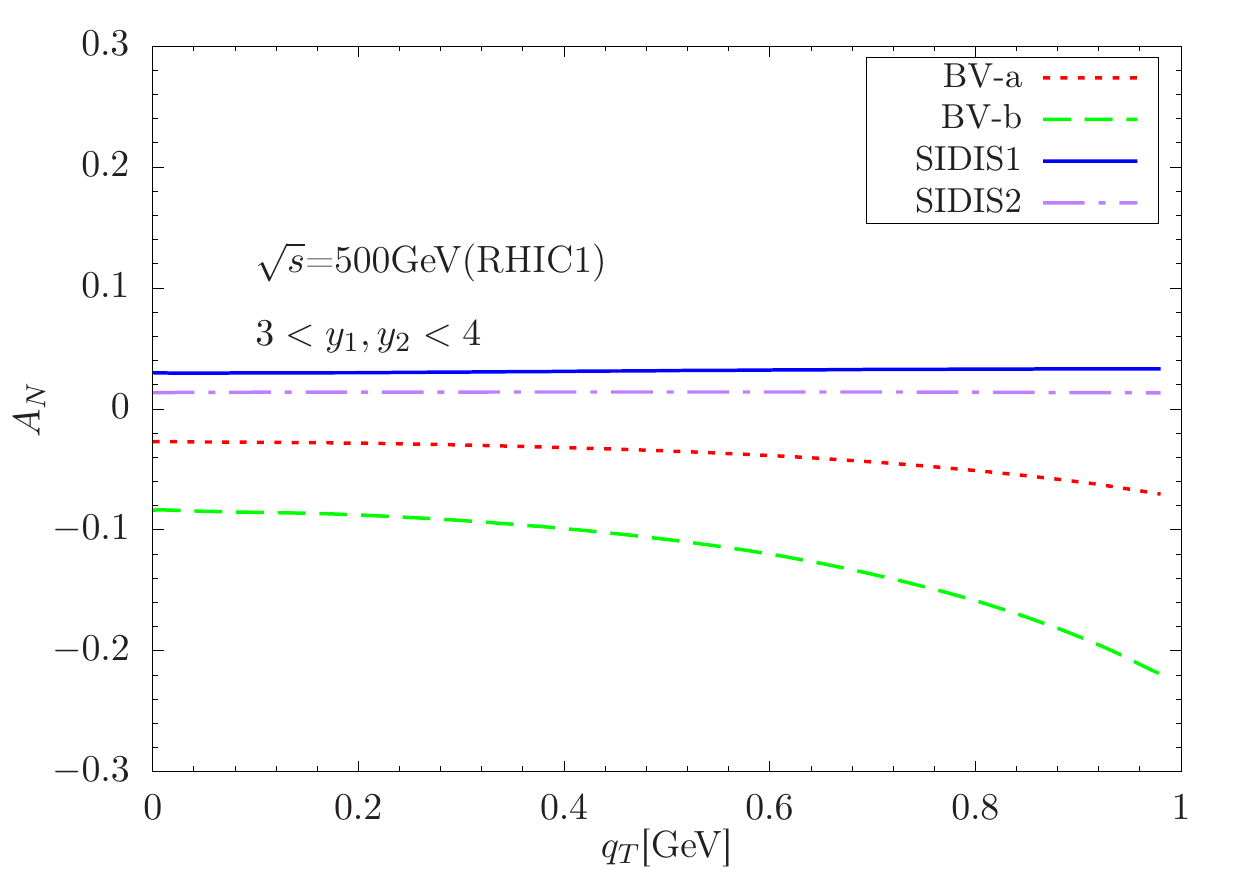}
\includegraphics[height=6.3cm,width=5.9cm,angle=0]{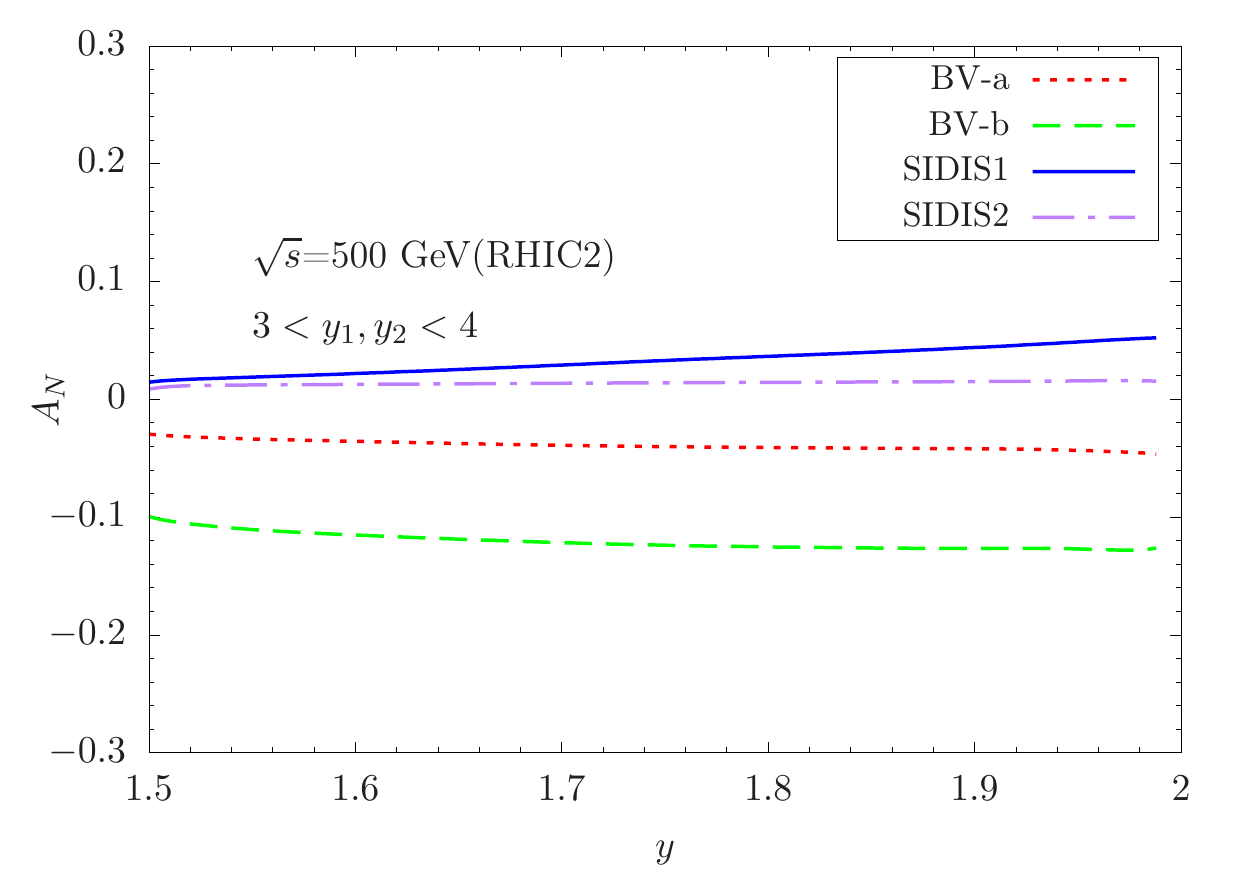}
\includegraphics[height=6.3cm,width=5.9cm,angle=0]{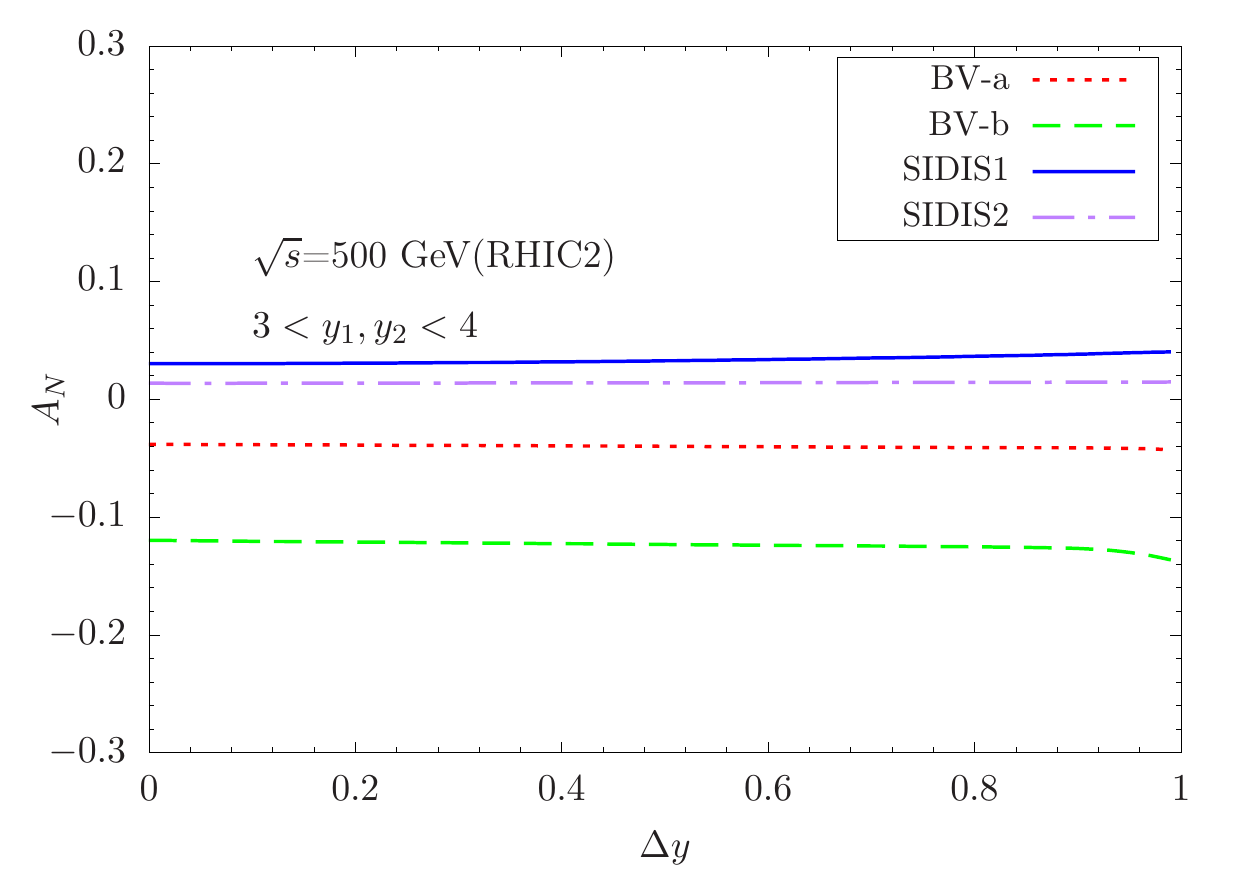}
\caption{\label{500_2}
 Single spin asymmetry in $\rm p+p^\uparrow \to J/\psi+J/\psi+X$ process as functions of $\rm q_T$, y, $\rm \Delta y$ with BV-a, BV-b, SIDIS1 and SIDIS2 parameters at $\rm \sqrt{s}=500$GeV (RHIC2). The integration ranges are $\rm 0 < q_T < 1$ GeV, and we impose cuts on both individual $\rm J/\psi$'s rapidity $\rm 3 < y_1,y_2 < 4$.}
\end{figure}

The results are shown in FIG.\ref{115}-\ref{500_2} and the conventions in these figures are the following. The "SIDIS1" and "SIDIS2" curves are obtained by using D'Alesio at el. \cite{DAlesio:2015fwo} fit parameters of GSF. The obtained asymmetry using Anselmino et al. \cite{Anselmino:2016uie} fit parameters is represented by "BV-a" and "BV-b".
As aforementioned, we have considered the final heavy quarks produced to be in the both CS and CO states for calculating the numerator and denominator part of Eq.(\ref{SSA}).
We find that the SSA with also color octet state contribution is very close to the one without color octet state contribution.

From FIG.\ref{115}-\ref{500_2}, the asymmetries are estimated 
to be positive and negative with SIDIS and BV parameterization respectively, as functions of $\rm q_T$, $\rm y$ and $\rm \Delta y$. The estimated asymmetry using "SIDIS2" fit is close to zero for all $\rm \sqrt{s}$ and the physical quantities $\rm q_T$, $\rm y$ and $\rm \Delta y$, considering our adopted cuts.
While the estimated $\rm |A_N|$ using "BV-b" fit is larger than all the other results in all CM energies and all considered physical quantities. In addition, the estimated asymmetries using all four parameters vary a little in the considered ranges of $\rm q_T$, $\rm y$ and $\rm \Delta y$. The obtained asymmetry using "BV-b" parameters is maximum about 12\% as functions of $\rm q_T$ and $\rm y$ at AFTER@LHC $\rm \sqrt{s}$ in FIG.\ref{115} (left panel) and FIG.\ref{115} (middle panel). It is also shown that the estimated asymmetry using "SIDIS1" fit is very close to the asymmetry using "SIDIS2" fit in AFTER@LHC and RHIC2 ($\rm 2<y_1,y_2<3$).
While in RHIC1 and RHIC2 ($\rm 3<y_1,y_2<4$) the "SIDIS1" fit results are larger than those from "SIDIS2". The sign of the asymmetry for "BV-a" and "BV-b" fit parameters depends on the magnitude of $\rm N_d$ and the relative magnitude of $\rm N_u$ and $\rm N_d$, and these have opposite sign which can be observed in TABLE.\ref{table1}. The negative sign magnitude of $\rm N_d$ is larger compared to $\rm N_u$ as a result the "BV-b" asymmetry is negative. That is to say, the modeling of GSF strongly decides the magnitude and sign of the asymmetry.

\begin{figure}[hbtp]
\includegraphics[height=6.3cm,width=5.9cm,angle=0]{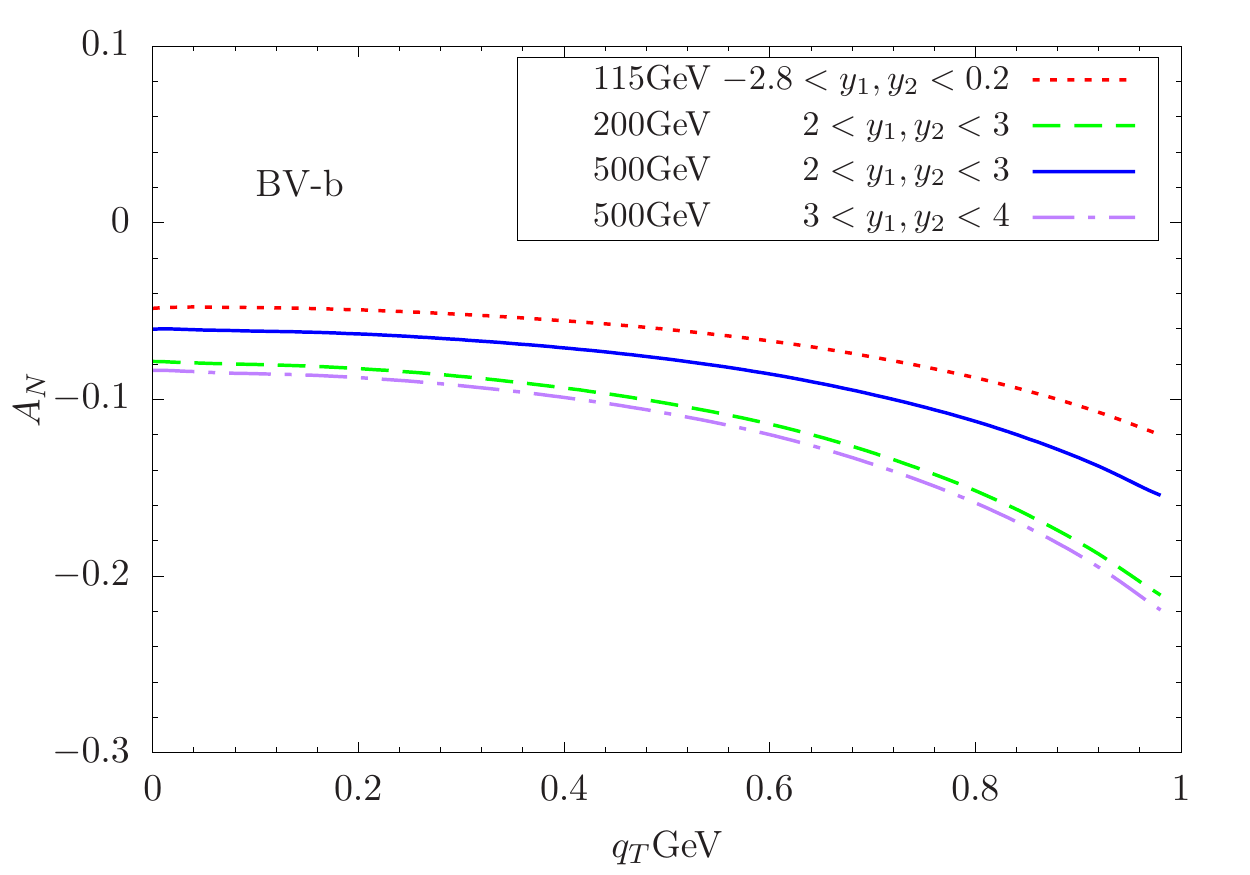}
\includegraphics[height=6.3cm,width=5.9cm,angle=0]{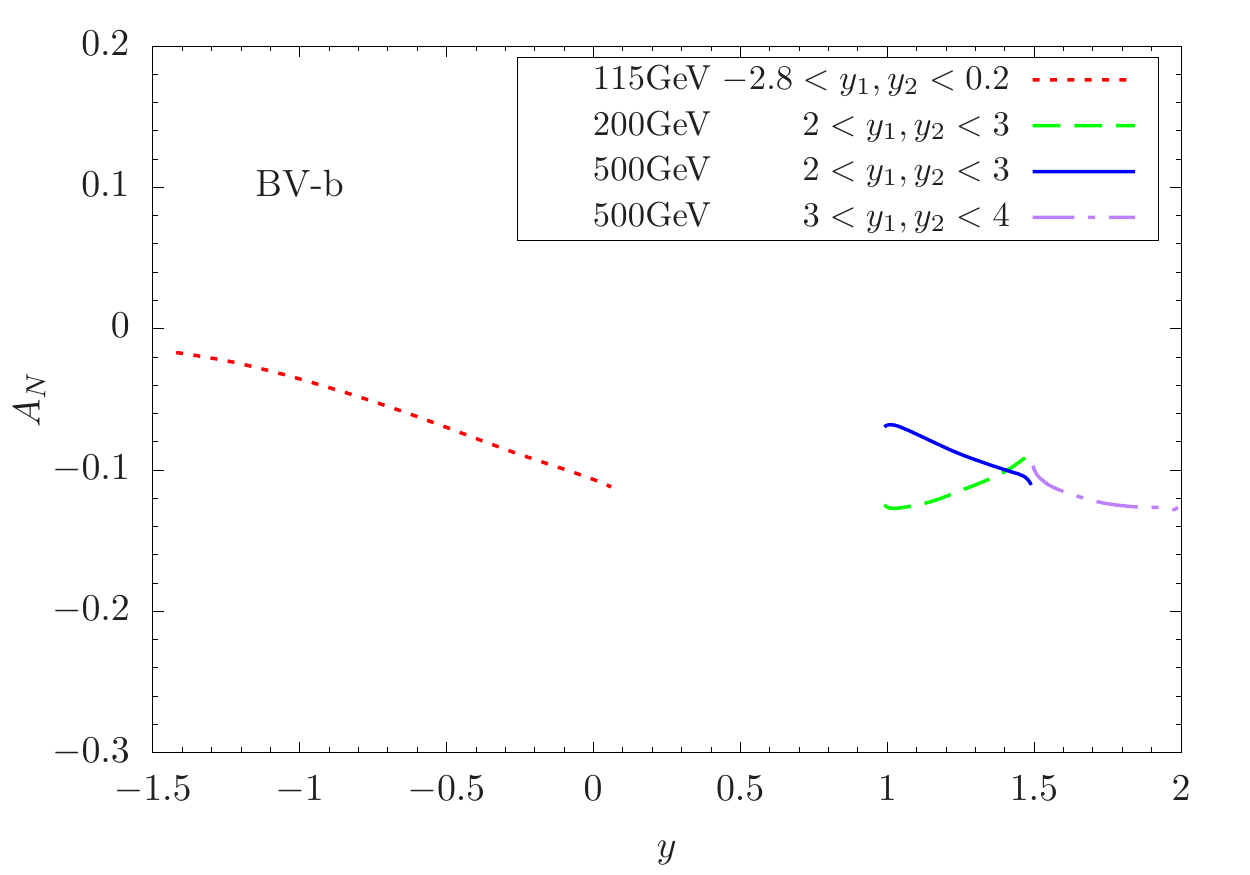}
\includegraphics[height=6.3cm,width=5.9cm,angle=0]{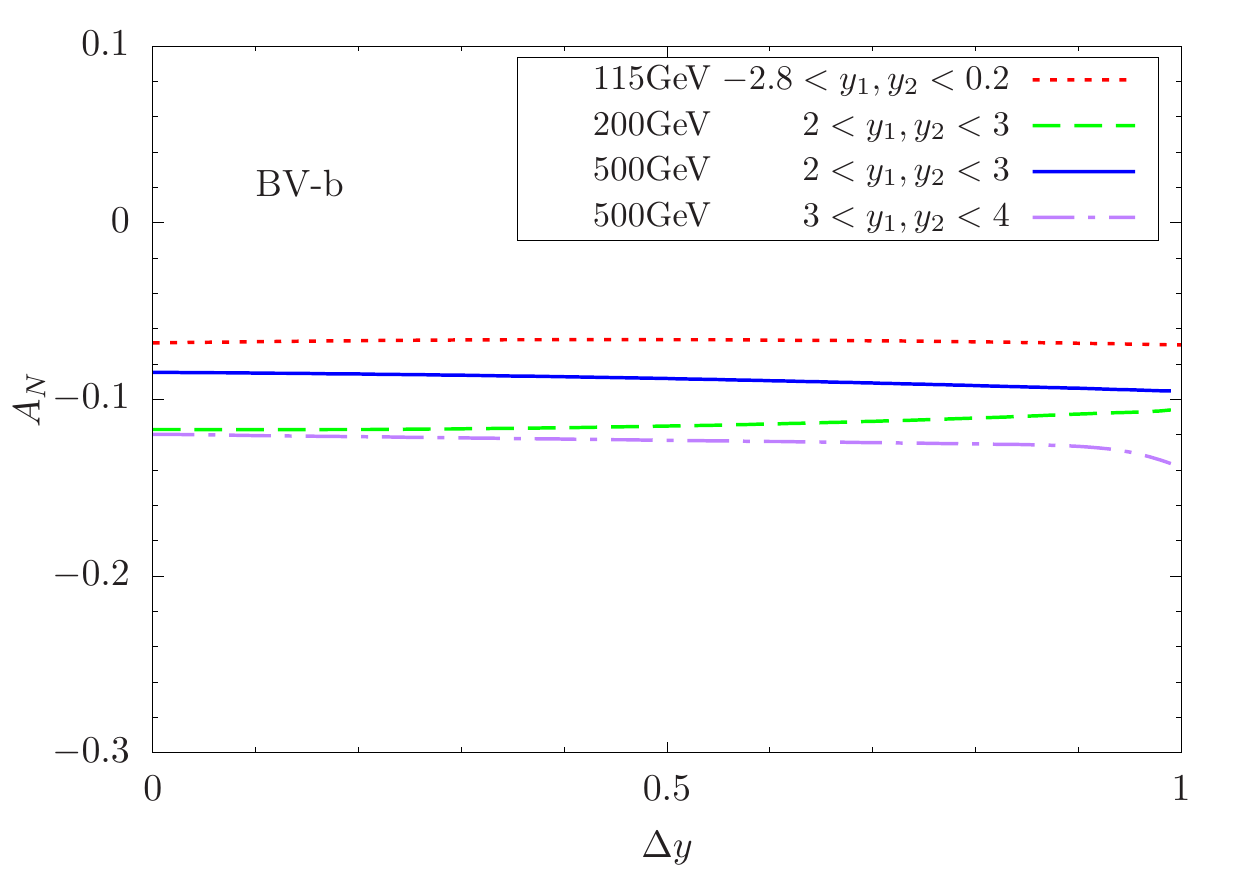}
\caption{\label{BVB}
 Predictions for asymmetry as functions of (the left panel) $\rm q_T$, (the middle panel) $y$ and (the right panel) $\rm \Delta y$ obtained using the BV-b GSF parameters for all the three CM energy values considered ($\rm \sqrt{s}=115$ GeV, 200 GeV, 500 GeV). The integration ranges are $\rm 0 < q_T < 1$ GeV, and we impose cuts on both individual $\rm J/\psi$'s rapidity $\rm -2.8 < y_1,y_2 < 0.2$ for $\rm \sqrt{s}=115$ GeV, $\rm 2< y_1,y_2 < 3$ and $\rm 3 < y_1,y_2 < 3.8$ for $\rm \sqrt{s}=200$ GeV, and $\rm 3 < y_1,y_2 < 4$ for $\rm \sqrt{s} = 500$ GeV. (Asymmetry peaks in negative y region for AFTER@LHC energy as we have used the convention for fix target experiments as explained in the Section \ref{numerical}). }
\end{figure}
\begin{figure}[hbtp]
\includegraphics[height=6.3cm,width=5.9cm,angle=0]{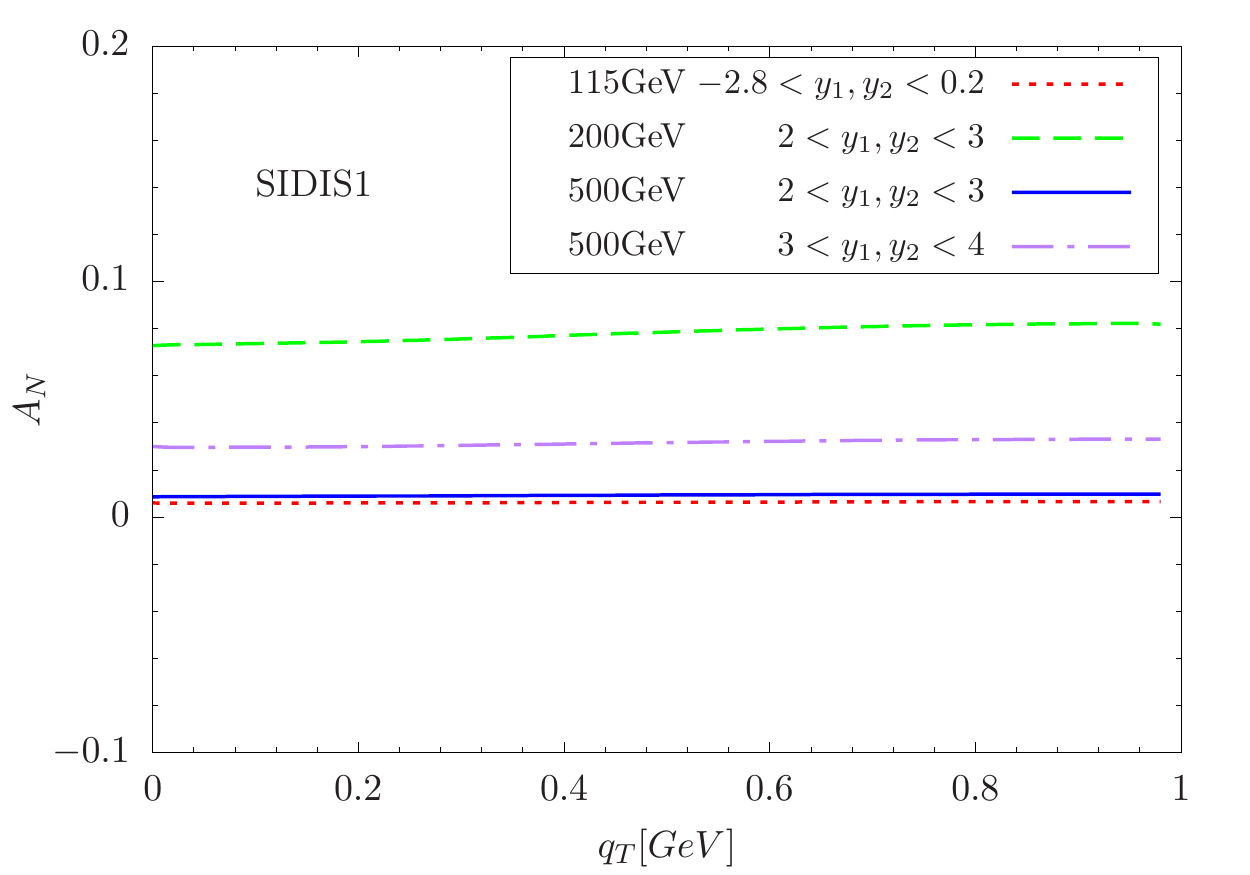}
\includegraphics[height=6.3cm,width=5.9cm,angle=0]{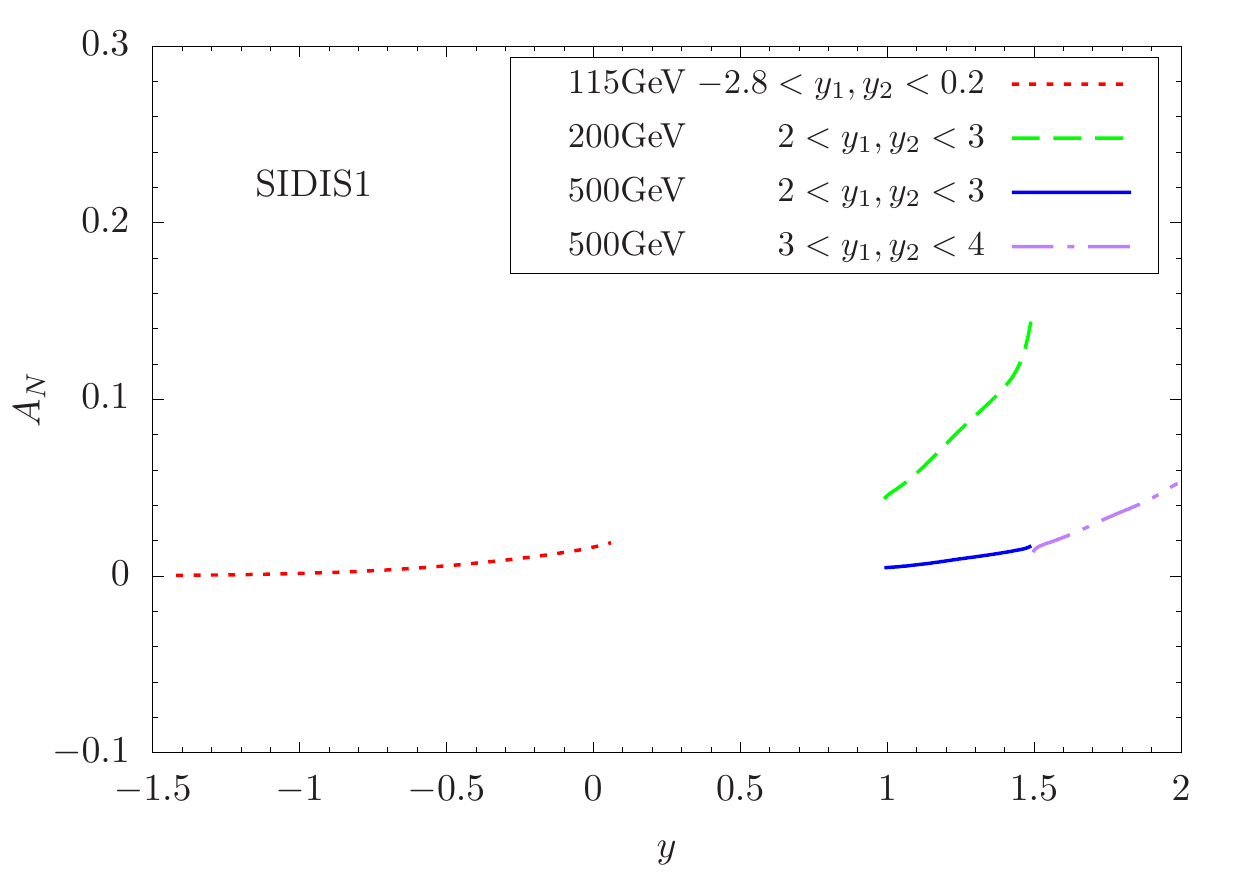}
\includegraphics[height=6.3cm,width=5.9cm,angle=0]{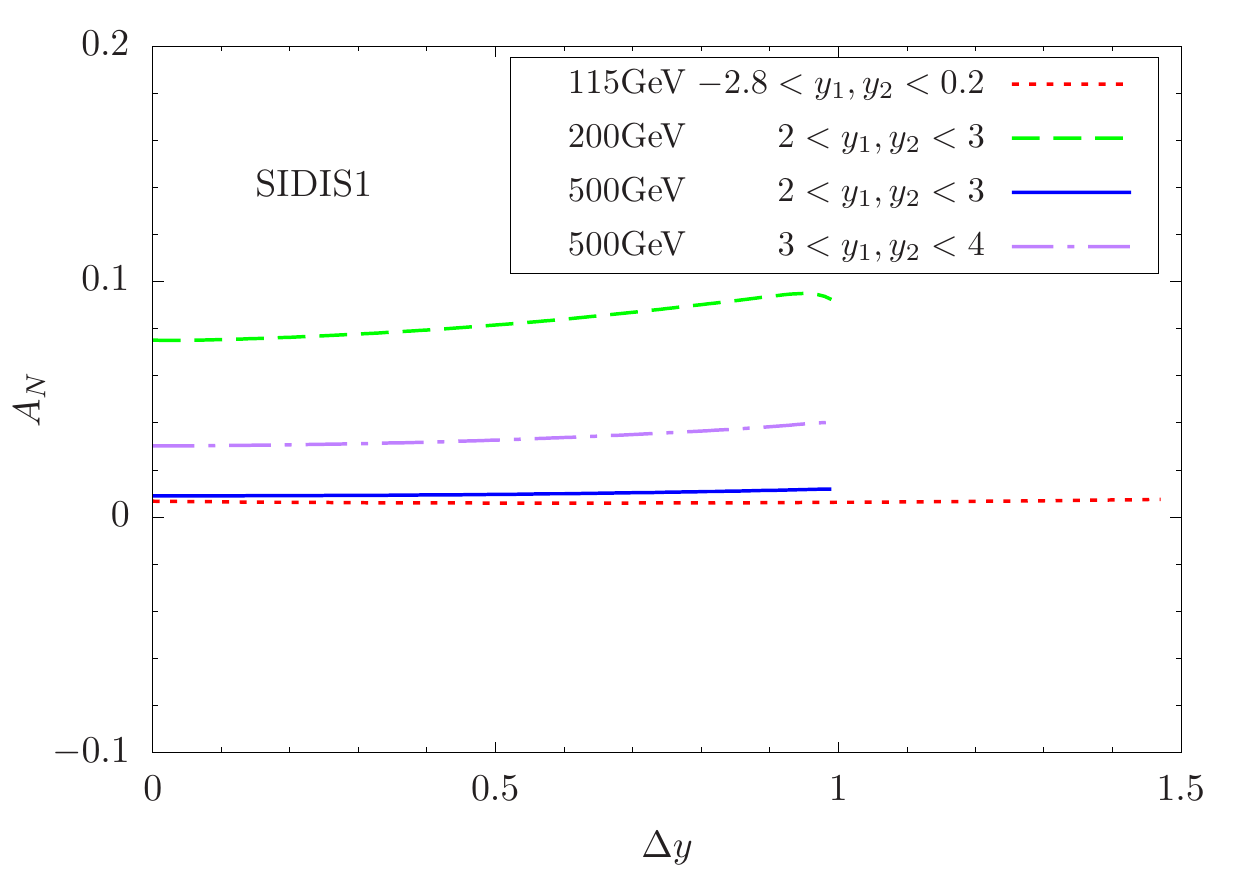}
\caption{\label{SIDIS1}
Predictions for SSA asymmetries as functions of (a) $\rm q_T$, ( the middle panel) $y$ and ( the right panel) $\rm \Delta y$ obtained using the SIDIS1 GSF parameters for all the three CM energy values considered ($\rm \sqrt{s}=115$ GeV, 200 GeV, 500 GeV). The integration ranges are $\rm 0 < q_T < 1$ GeV, and we impose cuts on both individual $\rm J/\psi$'s rapidity: $\rm -2.8 < y_1,y_2 < 0.2$ for $\rm \sqrt{s}=115$ GeV, $\rm 2< y_1,y_2 < 3$ and $\rm 3 < y_1,y_2 < 3.8$ for $\rm \sqrt{s}=200$ GeV, and $\rm 3 < y_1,y_2 < 4$ for $\rm \sqrt{s} = 500$ GeV. (Asymmetry peaks in negative y region for AFTER@LHC energy as we have used the convention for fix target experiments as explained in the Section \ref{numerical}).}
\end{figure}
In FIG.\ref{BVB} and Fig.\ref{SIDIS1}, we present the asymmetry predictions obtained with the two GSF fits, BV-b and SIDIS1, for all the three CM energies considered. 
It should be noted that in FIG.\ref{BVB} and \ref{SIDIS1}, the y distribution peaks are in negative region for AFTER@LHC CM energy. This is due to the fact that AFTER@LHC is a fixed target experiment and we have taken $\rm y_{cm}$ to be positive in the unpolarized beam direction. In FIG.\ref{BVB} and Fig.\ref{SIDIS1}, the y-distributions of the SSA lie from left to right as CM energy increases from 115 to 500 GeV, duing to the responding rapidity cuts. This is in contrast to RHIC1 and RHIC2 curves, where we have used the convention followed by PHENIX experiment, in which rapidity is considered to be positive in the forward hemisphere of the polarized proton. We find the largest asymmetry values are of about 20\% for the $\rm q_T$-distribution with BV-b fit and 11\% for the y-distribution with SIDIS1 fit. For the $\rm q_T$-distributions with BV-b fit, the asymmetry becomes larger as the value of $\rm q_T$ increases. Generally, the asymmetry increases when the CM energy decreases. Whereas, it is desired to notice in FIG.\ref{BVB} and \ref{SIDIS1} that the asymmetry for 115 GeV has the lowest value than all the others in $\rm q_T$, $\rm y$ and $\rm \Delta y$-distributions. This is because the energy increases and more strict rapidity cuts are applied, the unpolarized cross section decreases more quickly than the polarized one. Specially, one can find from FIG.\ref{SIDIS1} that 200 GeV (RHIC1) case gives the largest asymmetries in all three distributions. Moreover, in the case of $\rm q_T$-asymmetries, shown in FIG.\ref{BVB} (left panel) and \ref{SIDIS1} (left panel), we find that the functional form of the $\rm q_T$ dependence remains the same up to an overall factor that depends on $\rm \sqrt{s}$ and the rapidity range. This is also a reflection of the factorized $\rm q_T$ dependence that we have assumed for the TMDPDFs.

\section{SUMMARY AND DISCUSSIONS}
\label{summary}

In this paper, we have calculated the single spin asymmetry (SSA) in the double $\rm J/\psi$ production. Within the NRQCD framework, the color singlet state $\rm ^3 S_1^{(1)}$ and color octet state $\rm ^3 S_1^{(8)}$ contributions to the double $\rm J/\psi$ production are considered. We have given out the SSA as functions of $\rm p_T$, $\rm y$ and $\rm \Delta y$. Sizable asymmetry is obtained as functions of them in the kinematic range $\rm \hat{s}=M^2 \gg \Lambda_{QCD}$, $\rm 0 < q_T < 1\ \text{GeV}$, $\rm 4 < p_{2T} < 6\ \text{GeV}$. Typically we have considered $\rm -2.8 < y_1,y_2 < 0.2$ for $\rm \sqrt{s}=115$ GeV, $\rm 2< y_1,y_2 < 3$ and $\rm 3 < y_1,y_2 < 3.8$ for $\rm \sqrt{s}=200$ GeV, and $\rm 3 < y_1,y_2 < 4$ for $\rm \sqrt{s} = 500$ GeV. We find the asymmetry can reach more than 10 percent using BV-b fit. The results using the other three fit parameters are also not small. The sizable asymmetry indicates that the double $\rm J/\psi$ production in proton proton collision is a competitive process to probe the gluon Sivers function over a wide kinematic region accessible at the RHIC and AFTER@LHC.

\begin{acknowledgments}
Xuan Luo thanks professor Sergey Baranov and professor Asmita Mukherjee for very useful discussions. Hao Sun is supported by the National Natural Science Foundation of China (Grant No.11675033), and by the Fundamental Research Funds for the Central Universities (Grant No. DUT18LK27).
\end{acknowledgments}

\appendix
\section{S\lowercase{quare of the amplitude for} $\rm g+ g\rightarrow J/\psi+J/\psi$ \lowercase{process}}
\label{ap1}

The amplitude squares of gluon-gluon to $\rm J/\psi$ pair can be calculated by the FORM package \cite{Kuipers:2012rf} straightforwardly and we have cross checked with paper \cite{Li:2009ug}\cite{Qiao:2002rh}. The amplitude squares of $\rm ^3 S_1^{(1)}$, $\rm ^3 S_1^{(1)}$ and $\rm ^3 S_1^{(8)}$, $\rm ^3 S_1^{(8)}$ are given below
\begin{equation}\label{d4}
\begin{aligned}
|\mathcal{\overline{M}}[^3 S_1^{(1)},^3 S_1^{(1)}]|^2 = 
& \frac{256\alpha_s^4 \pi^2 |R(0)|^4}{81 m^2 \hat{s}^6 {(m^2-\hat{t})}^4 {(m^2-\hat{u})}^4} [7776 m^{24}-31536 m^{22} \hat{s}-93312 m^{22}
\hat{u}+619979374 m^{20} \hat{s}^2\\
& +362016 m^{20}
\hat{s} \hat{u}+513216 m^{20} \hat{u}^2-1549870224
m^{18} \hat{s}^3-1550415136 m^{18} \hat{s}^2
\hat{u}\\
& -1885680 m^{18} \hat{s} \hat{u}^2-1710720 m^{18}
\hat{u}^3 +1549859868 m^{16} \hat{s}^4+3100230672
m^{16} \hat{s}^3 \hat{u}\\
& +1552794856 m^{16} \hat{s}^2
\hat{u}^2+5883840 m^{16} \hat{s} \hat{u}^3+3849120
m^{16} \hat{u}^4-774937124 m^{14}
\hat{s}^5\\
& -2325156672 m^{14} \hat{s}^4
\hat{u}-2327516108 m^{14} \hat{s}^3 \hat{u}^2-783601712
m^{14} \hat{s}^2 \hat{u}^3-12221280 m^{14} \hat{s}
\hat{u}^4\\
& -6158592 m^{14} \hat{u}^5+193740424
m^{12} \hat{s}^6+775141866 m^{12} \hat{s}^5
\hat{u}+1164153206 m^{12} \hat{s}^4 \hat{u}^2 \\
& +782390040
m^{12} \hat{s}^3 \hat{u}^3+210283340 m^{12} \hat{s}^2
\hat{u}^4+17744832 m^{12} \hat{s} \hat{u}^5+7185024
m^{12} \hat{u}^6\\
& -19376450 m^{10}
\hat{s}^7-96950056 m^{10} \hat{s}^6 \hat{u}-194537756
m^{10} \hat{s}^5 \hat{u}^2-197953656 m^{10} \hat{s}^4
\hat{u}^3\\
& -109651440 m^{10} \hat{s}^3 \hat{u}^4-40897820
m^{10} \hat{s}^2 \hat{u}^5-18379872 m^{10} \hat{s}
\hat{u}^6-6158592 m^{10} \hat{u}^7\\
& +587 m^8 \hat{s}^8+19710 m^8 \hat{s}^7 \hat{u}+244772 m^8 \hat{s}^6
\hat{u}^2+1603468 m^8 \hat{s}^5 \hat{u}^3+6229962 m^8
\hat{s}^4 \hat{u}^4\\
& +14478304 m^8 \hat{s}^3 \hat{u}^5+19359816
m^8 \hat{s}^2 \hat{u}^6+13582080 m^8 \hat{s}
\hat{u}^7+3849120 m^8 \hat{u}^8-40 m^6 \hat{s}^9\\
& -2370
m^6 \hat{s}^8 \hat{u}-44306 m^6 \hat{s}^7 \hat{u}^2-387560
m^6 \hat{s}^6 \hat{u}^3-1930716 m^6 \hat{s}^5
\hat{u}^4-5856736 m^6 \hat{s}^4 \hat{u}^5\\
& -10863572 m^6
\hat{s}^3 \hat{u}^6-11899056 m^6 \hat{s}^2 \hat{u}^7-7017840
m^6 \hat{s} \hat{u}^8-1710720 m^6 \hat{u}^9+m^4
\hat{s}^{10}\\
& +76 m^4 \hat{s}^9 \hat{u}+3756 m^4 \hat{s}^8
\hat{u}^2+52062 m^4 \hat{s}^7 \hat{u}^3+353472 m^4 \hat{s}^6
\hat{u}^4+1398834 m^4 \hat{s}^5 \hat{u}^5\\
& +3421754 m^4
\hat{s}^4 \hat{u}^6+5210968 m^4 \hat{s}^3 \hat{u}^7+4784622
m^4 \hat{s}^2 \hat{u}^8+2414880 m^4 \hat{s} \hat{u}^9\\
& +513216 m^4 \hat{u}^{10}-36 m^2 \hat{s}^9 \hat{u}^2-2668 m^2
\hat{s}^8 \hat{u}^3-31068 m^2 \hat{s}^7 \hat{u}^4-172796 m^2
\hat{s}^6 \hat{u}^5\\
& -560620 m^2 \hat{s}^5 \hat{u}^6-1134624 m^2 \hat{s}^4 \hat{u}^7-1450460 m^2 \hat{s}^3
\hat{u}^8-1136880 m^2 \hat{s}^2 \hat{u}^9\\
& -498096 m^2 \hat{s} \hat{u}^{10}-93312 m^2 \hat{u}^{11}+698 \hat{s}^8
\hat{u}^4	+7400 \hat{s}^7 \hat{u}^5+35004 \hat{s}^6 \hat{u}^6+95528
\hat{s}^5 \hat{u}^7\\
& +163418 \hat{s}^4 \hat{u}^8+178560 \hat{s}^3
\hat{u}^9+121248 \hat{s}^2 \hat{u}^{10}+46656 \hat{s}
\hat{u}^{11}+7776 \hat{u}^{12}	]
\end{aligned}
\end{equation}
\begin{equation}\label{d4}
\begin{aligned}
|\mathcal{\overline{M}}[^3 S_1^{(8)},^3 S_1^{(8)}]|^2 = 
& \frac{4\pi^4 \alpha_s^4 \langle O_8(^3 S_1) \rangle^2}{243m^6\hat{s}^6(t-m^2)^4(u-m^2)^4}[187272 m^{28}-72 m^{26} (11537 \hat{s}+31194
   \hat{t})+6 m^{24} (259913 \hat{s}^2
\\
&+1570956 \hat{s} \hat{t}+2057724 \hat{t}^2)
-8 m^{22} (233734 \hat{s}^3+2111409 \hat{s}^2
   \hat{t}+6071274 \hat{s} \hat{t}^2+5141880 \hat{t}^3)
\\
&+4 m^{20} (446021 \hat{s}^4+4708219 \hat{s}^3
   \hat{t}+20480415 \hat{s}^2 \hat{t}^2+37508616 \hat{s}
   \hat{t}^3+23128740 \hat{t}^4)
\\
&-2 m^{18} (674202 \hat{s}^5+7783209 \hat{s}^4
   \hat{t}+41993932 \hat{s}^3 \hat{t}^2+117212424 \hat{s}^2
   \hat{t}^3+154359000 \hat{s} \hat{t}^4
\\
&+73984752 \hat{t}^5)+m^{16} (775181
   \hat{s}^6+9628777 \hat{s}^5 \hat{t}+60464369 \hat{s}^4
   \hat{t}^2+217547464 \hat{s}^3 \hat{t}^3
\\
&+438545220 \hat{s}^2 \hat{t}^4+444319344 \hat{s} \hat{t}^5+172576656
   \hat{t}^6)-2 m^{14} (158802
   \hat{s}^7+2143917 \hat{s}^6 \hat{t}
\\
&+15477603 \hat{s}^5 \hat{t}^2+67698320 \hat{s}^4 \hat{t}^3+180289870 \hat{s}^3
   \hat{t}^4+280325328 \hat{s}^2 \hat{t}^5+228221280 \hat{s}
   \hat{t}^6
\\
&+73941984 \hat{t}^7)+2 m^{12} (43072 \hat{s}^8+638490 \hat{s}^7 \hat{t}+5393635
   \hat{s}^6 \hat{t}^2+28486982 \hat{s}^5 \hat{t}^3
\\
&+94986651 \hat{s}^4
   \hat{t}^4+198281780 \hat{s}^3 \hat{t}^5+248119176 \hat{s}^2
   \hat{t}^6+167349456 \hat{s} \hat{t}^7+46204020
   \hat{t}^8)
\\
&-2 m^{10} (8039
   \hat{s}^9+112887 \hat{s}^8 \hat{t}+1157014 \hat{s}^7
   \hat{t}^2+7632256 \hat{s}^6 \hat{t}^3+31876569 \hat{s}^5
   \hat{t}^4
\\
&+85147430 \hat{s}^4 \hat{t}^5+144700858 \hat{s}^3
   \hat{t}^6+150182520 \hat{s}^2 \hat{t}^7+85844772 \hat{s}
   \hat{t}^8+20531880 \hat{t}^9)
\\
&+2 m^8
   (935 \hat{s}^{10}+9398 \hat{s}^9 \hat{t}+117747
   \hat{s}^8 \hat{t}^2+1103652 \hat{s}^7 \hat{t}^3+6182220 \hat{s}^6
   \hat{t}^4+21423546 \hat{s}^5 \hat{t}^5
\\
&+47491450 \hat{s}^4
   \hat{t}^6+67574132 \hat{s}^3 \hat{t}^7+59508939 \hat{s}^2
   \hat{t}^8+29339460 \hat{s} \hat{t}^9+6158916
   \hat{t}^{10})
\\
&-2 m^6 \hat{t} (57
   \hat{s}^{10}+1233 \hat{s}^9 \hat{t}+46541 \hat{s}^8
   \hat{t}^2+513120 \hat{s}^7 \hat{t}^3+2646793 \hat{s}^6
   \hat{t}^4+7942109 \hat{s}^5 \hat{t}^5
\\
&+15041136 \hat{s}^4
   \hat{t}^6+18324922 \hat{s}^3 \hat{t}^7+13942380 \hat{s}^2
   \hat{t}^8+6013080 \hat{s} \hat{t}^9+1119744
   \hat{t}^{10})
\\
&+m^4 \hat{t} (243
   \hat{s}^{11}+3951 \hat{s}^{10} \hat{t}+6714 \hat{s}^9
   \hat{t}^2+14420 \hat{s}^8 \hat{t}^3+179582 \hat{s}^7
   \hat{t}^4+919446 \hat{s}^6 \hat{t}^5
\\
&+2488136 \hat{s}^5
   \hat{t}^6+4132862 \hat{s}^4 \hat{t}^7+4395900 \hat{s}^3
   \hat{t}^8+2933988 \hat{s}^2 \hat{t}^9+1119744 \hat{s}
   \hat{t}^{10}
\\
&+186624 \hat{t}^{11})-162 m^2
   \hat{s}^3 \hat{t}^2 (\hat{s}+\hat{t}) (\hat{s}^2+\hat{s}
   \hat{t}+\hat{t}^2) (9 \hat{s}^5+19 \hat{s}^4 \hat{t}+20
   \hat{s}^3 \hat{t}^2+7 \hat{s}^2 \hat{t}^3-3 \hat{s} \hat{t}^4
\\
&-6 \hat{t}^5)+243 \hat{s}^4 \hat{t}^2 (\hat{s}+\hat{t})^2
   (\hat{s}^2+\hat{s} \hat{t}+\hat{t}^2)^3]
\end{aligned}
\end{equation}
where m is the mass of $\rm J/\psi$; $\rm |R(0)|$ is the magnitude of its radial wave function at origin; $\rm \hat{s}$, $\rm \hat{t}$, $\rm \hat{u}$ are the usual Mandelstam variables.

\bibliography{a}

\end{document}